\documentclass[tighten]{emulateapj}
\usepackage{lscape}
\usepackage{amsmath}
\tabletypesize{\scriptsize}

\newcommand{\halpha}{H$\alpha$}

\shorttitle{Star Formation in Bulgeless Disk Galaxies}

\begin{document}

\journalinfo{Accepted for Publication in {\it The Astrophysical
Journal}}

\title{Properties of Bulgeless Disk Galaxies II. Star Formation as a
Function of Circular Velocity}

\author{Linda~C.~Watson\altaffilmark{1,2},
  Paul~Martini\altaffilmark{2,3}, Ute~Lisenfeld\altaffilmark{4},
  Man-Hong Wong\altaffilmark{2,5}, Torsten~B\"oker\altaffilmark{6},
  Eva~Schinnerer\altaffilmark{7}}

\altaffiltext{1}{Harvard-Smithsonian Center for Astrophysics, 60
Garden Street, Cambridge, MA 02138, USA; lwatson@cfa.harvard.edu}
\altaffiltext{2}{Department of Astronomy, The Ohio State University,
  140 West 18th Avenue, Columbus, OH 43210, USA}
\altaffiltext{3}{Center for Cosmology and AstroParticle Physics, The
  Ohio State University, 191 West Woodruff Avenue, Columbus, OH 43210,
  USA}
\altaffiltext{4}{Departamento de F\'isica Te\'orica y del Cosmos, Universidad
  de Granada, Spain}
\altaffiltext{5}{Department of Physics, University of Illinois, 1110
West Green Street, Urbana, IL 61801, USA}
\altaffiltext{6}{European Space Agency, Keplerlaan 1, 2200 AG
  Noordwijk, The Netherlands}
\altaffiltext{7}{Max-Planck-Institut f\"ur Astronomie, K\"onigstuhl 17,
  D-69117 Heidelberg, Germany}

\begin{abstract}
We study the relation between the surface density of gas and star
formation rate in twenty moderately-inclined, bulgeless disk galaxies
(Sd-Sdm Hubble types) using CO(1--0) data from the IRAM $30 \, {\rm
m}$ telescope, \ion{H}{1} emission line data from the VLA/EVLA,
H$\alpha$ data from the MDM Observatory, and PAH emission data derived
from {\it Spitzer} IRAC observations.  We specifically investigate the
efficiency of star formation as a function of circular velocity
($v_{\rm circ}$).  Previous work found that the vertical dust
structure and disk stability of edge-on, bulgeless disk galaxies
transition from diffuse dust lanes with large scale heights and
gravitationally-stable disks at $v_{\rm circ} < 120 \, {\rm km \,
s^{-1}}$ ($M_{\ast} \lesssim 10^{10} \, M_{\odot}$) to narrow dust
lanes with small scale heights and gravitationally-unstable disks at
$v_{\rm circ} > 120 \, {\rm km \, s^{-1}}$.  We find no transition in
star formation efficiency ($\Sigma_{\rm SFR}/\Sigma_{\rm HI+H_{2}}$)
at $v_{\rm circ} = 120 \, {\rm km \, s^{-1}}$, or at any other
circular velocity probed by our sample ($v_{\rm circ} = 46 - 190 \,
{\rm km \, s^{-1}}$).  Contrary to previous work, we find no
transition in disk stability at any circular velocity in our sample.
Assuming our sample has the same dust structure transition as the
edge-on sample, our results demonstrate that scale height differences
in the cold interstellar medium of bulgeless disk galaxies do not
significantly affect the molecular fraction or star formation
efficiency.  This may indicate that star formation is primarily
affected by physical processes that act on smaller scales than the
dust scale height, which lends support to local star formation models.
\end{abstract}


\section{Introduction}
\label{sec:intro}
High resolution studies of the Milky Way and nearby galaxies show that
star formation mainly occurs in giant molecular clouds (GMCs), with
typical GMC masses between about $10^{3}$ and $10^{7} \, M_{\odot}$
and radii between about ten and several hundred parsecs
\citep[e.g.,][]{fukui10}.  It is challenging to study molecular gas on
these scales in even the nearest galaxies \citep[$\lesssim 4 \, {\rm
Mpc}$; e.g.,][]{bolatto08} and studies of more distant galaxies can
only currently investigate the average properties of gas and stars on
larger scales.  An advantage of more distant galaxies is that they
exhibit a much wider range of physical properties, which enables
investigation into the influence of environment (e.g., mid-plane
pressure, metallicity, and scale height) and processes that act on
large scales (e.g., shear and large-scale gravitational instabilities)
on star formation.

Extragalactic studies of star formation have found that the star
formation rate (SFR) surface density ($\Sigma_{\rm SFR}$) and gas
surface density ($\Sigma_{\rm gas}$) are correlated in the form of the
Kennicutt-Schmidt law: $\Sigma_{\rm SFR} \propto \Sigma_{\rm gas}^{N}$
\citep{schmidt59, kennicutt98}.  This star formation law has been
studied by averaging over scales as small as about $100 \, {\rm pc}$
and as large as the entire optical disk.  Recent high-resolution
studies have found that the SFR surface density is more strongly
correlated with the molecular gas surface density ($\Sigma_{\rm
H_{2}}$) than with the atomic gas surface density ($\Sigma_{\rm HI}$),
with $\Sigma_{\rm SFR} \propto \Sigma_{\rm H_{2}}^{N}$ and $N$ between
0.8 and 1.5
\citep{wong02,kennicutt07,bigiel08,blanc09,schruba10,liu11,rahman11}.
This result confirms, from an extragalactic perspective, that stars
form from molecular gas and has led to an expansion in the scope of
many star formation studies to investigate how environment and
large-scale processes affect the molecular fraction in the
interstellar medium (ISM).

\citet{leroy08} addressed environmental effects on the molecular
fraction and star formation efficiency (SFE) with high-quality, $750
\, {\rm pc}$-resolution observations of $\Sigma_{\rm SFR}$,
$\Sigma_{\rm HI}$, and $\Sigma_{\rm H_{2}}$ over the optical disk of
23 nearby galaxies.  They compared these observations to many star
formation models and thresholds and concluded that no model fit the
data sufficiently well to be declared a clear favorite.  This result
led them to suggest that physics that acts on scales smaller than
their resolution is most important for setting the molecular fraction
and SFE.

While no model was an ideal fit to the data presented in
\citet{leroy08}, the best fit was arguably with a model in which the
ratio of molecular to atomic surface density is related to the
mid-plane pressure ($P_{\rm h}$): $R_{\rm mol} \equiv \Sigma_{\rm
H_{2}}/\Sigma_{\rm HI} \propto P_{\rm h}^{\alpha}$, and the molecular
SFE (${\rm SFE[H_{2}}] \equiv \Sigma_{\rm SFR}/\Sigma_{\rm H_{2}}$) is
constant, such that ${\rm SFE} \equiv \Sigma_{\rm SFR}/\Sigma_{\rm
HI+H_{2}} = {\rm SFE({\rm H_{2}})} \, \frac{R_{\rm mol}}{R_{\rm
mol}+1}$.  \citet{leroy08} noted that a model where the molecular SFE
is constant requires the population of GMCs in any region to be
sampled from the same distribution of properties (e.g., size and
mass), independent of environment.  Furthermore, once GMCs form, the
general environment cannot have a strong influence on their
properties.  Finally, one must compare the model to observations with
a number of GMCs per resolution element so as to average over
evolutionary effects.  \citet{elmegreen93} predicted that $R_{\rm
mol}$ should depend on the mid-plane pressure and the interstellar
radiation field ($j$): $R_{\rm mol} \propto P_{\rm h}^{2.2} \,
j^{-1}$.  Assuming $\Sigma_{\rm SFR} \propto \Sigma_{\rm H_{2}}$ and
$j \propto \Sigma_{\rm SFR}$, the model predicts $R_{\rm mol} \propto
P_{\rm h}^{\alpha}$, with $\alpha = 1.2$.  \citet{wong02} studied
seven molecular-dominated spiral galaxies and found $\alpha=0.8$,
while \citet{blitz06} found $\alpha=0.92$ in their study of fourteen
galaxies with a large range of $R_{\rm mol}$. Values between $\alpha =
0.5$ and 1.2 encompass most of the \citet{leroy08} data on the SFE
versus $P_{\rm h}$ plane.

\citet{ostriker10} recently presented a star formation model that
produces an approximately linear relationship between $R_{\rm mol}$
and $P_{\rm h}$.  The authors divided the ISM into a diffuse component
and a gravitationally bound component.  The fraction of gas in each
component is set by the requirement that gas pressure in the diffuse
component is balanced by the gravity of stars, dark matter, and gas
(both diffuse and bound), while heating (mainly by ultraviolet (UV)
photons from O and B stars formed in the bound component) balances
cooling.  The model assumes that the SFE within the gravitationally
bound component is constant.  \citet{bolatto11} added a
metallicity-dependent heating term to the \citet{ostriker10} model,
which resulted in agreement between the large \ion{H}{1} surface
densities observed in the Small Magellanic Cloud and the surface
density of diffuse gas calculated with the adjusted \citet{ostriker10}
model.

Another leading star formation model is that of \citet{krumholz09},
where the molecular fraction is determined by processes that act on
scales no larger than $\sim100 \, {\rm pc}$, which is the size of
atomic-molecular complexes in the model.  Specifically, the molecular
fraction is set by the balance between the formation of molecular
hydrogen on the surfaces of dust grains and the destruction of
molecular hydrogen by UV photons.  Both dust shielding and ${\rm
H_{2}}$ self-shielding contribute to the survival of molecular
hydrogen in the interior of the complexes.  Stars form only from
molecular gas and the model produces a constant molecular SFE because
properties of molecular clouds, like $\Sigma_{\rm H_{2}}$, are
independent of general ISM conditions, at least while gas surface
densities in the general ISM are less than GMC densities ($\sim 85 \,
M_{\odot} \, {\rm pc}^{-2}$).

In this paper, we study star formation in a sample of 20 bulgeless
disk galaxies.  Bulgeless galaxies are interesting from a number of
perspectives.  Their existence in relatively large numbers
\citep[$\gtrsim$15\% of disk galaxies;][]{kautsch06, kormendy10}
provides an important constraint on hierarchical galaxy formation
models, in which galaxies generally have rich merger histories that
lead to bulge growth.  The agreement between models and observations
is becoming better as feedback, high gas fractions, and satellite
mergers on radial orbits are included in the models (Robertson et
al. 2006, Hopkins et al. 2008, Brook et al. 2011; but see also
Scannapieco et al. 2011).  Under the assumption that bulges do form
when significant merger events occur, bulgeless galaxies are a
suitable sample to study secular evolution, where internal processes
like star formation lead to changes in the galaxies, such as bulge
growth \citep{kormendy04}

A number of works have studied the components of star formation in
late-type disk galaxies.  \citet{boeker03} found that their sample of
47 late-type spirals are similar to earlier-type spirals in that they
fall on the approximately linear correlation between far-infrared
(FIR) luminosity, which traces star formation, and the molecular
hydrogen mass within the central few kpc.  Furthermore,
\citet{matthews05} found that low-surface brightness, late-type disks
lie on this same relation.  \citet{dalcanton04} studied the dust and
cold ISM structure in a sample of 49 edge-on bulgeless disk galaxies.
They inferred that there is a sharp transition in dust lane structure
with circular velocity ($v_{\rm circ}$) based on measurements of $R-K$
color versus height above the midplane.  Galaxies with $v_{\rm circ} <
120 \, {\rm km \, s^{-1}}$ (we also refer to these as low-$v_{\rm
circ}$ galaxies) appear to have no dust lanes while galaxies with
$v_{\rm circ} > 120 \, {\rm km \, s^{-1}}$ (high-$v_{\rm circ}$
galaxies) have well-defined dust lanes \citep[the stellar mass
Tully-Fisher relation relates $v_{\rm circ} = 120 \, {\rm km \,
s^{-1}}$ with $M_{\ast} \sim 10^{10} \, M_{\odot}$;][]{bell01}.  The
authors concluded that the transition is likely due to a transition in
dust scale height rather than due to a sharp transition in the
quantity of dust present: low-$v_{\rm circ}$ galaxies have diffuse
dust lanes with large scale heights while high-$v_{\rm circ}$ galaxies
have dust lanes with smaller scale heights.  They came to this
conclusion because the dust structure transition occurs over a
relatively narrow range in circular velocity ($\sim 10 \, {\rm km \,
s^{-1}}$), and therefore over a relatively narrow range in gas and
total mass, where the dust-to-gas ratio (DGR) does not vary
substantially.

\citet{dalcanton04} also found that disk stability, parametrized by a
generalized Toomre Q parameter including both gas and stars
\citep{rafikov01}, is correlated with the dust structure, with
low-$v_{\rm circ}$ galaxies generally stable and high-$v_{\rm circ}$
galaxies generally unstable.  Furthermore, they concluded that a sharp
change in the contribution of turbulence to the stability parameter is
the likely cause of the stability and dust scale height transitions.
The authors suggested that high-$v_{\rm circ}$ galaxies have
turbulence dominated by supernovae explosions and gravitational
instabilities while low-$v_{\rm circ}$ galaxies have turbulence
dominated by only supernovae.  Independent of the source of the
turbulence, the turbulent velocities must be lower in the
high-$v_{\rm circ}$ galaxies to explain the stability results.  An
alternative interpretation for the dust structure transition is that
it is due to differences in stellar surface density and dust opacity,
as suggested by \citet{hunter06}.

\citet{dalcanton04} noted that the cold, star-forming gas should have
a similar distribution to the dust, with larger scale heights in
low-$v_{\rm circ}$ galaxies compared to high-$v_{\rm circ}$ galaxies.
They hypothesized that a transition in SFE might accompany the
transition in dust scale height and stability if the volume density of
gas is the relevant quantity for setting the SFR.  A low-$v_{\rm
circ}$ galaxy with a larger scale height but the same $\Sigma_{\rm
gas}$ relative to a high-$v_{\rm circ}$ galaxy will have a lower gas
volume density ($\rho_{\rm gas}$).  The low-$v_{\rm circ}$ galaxy
likely also has a lower gas pressure because pressure is proportional
to the gas volume density ($P \propto \rho_{\rm gas} \, \sigma_{\rm
gas}^2$, where $\sigma_{\rm gas}$ is the gas velocity dispersion).  In
the context of the star formation model where the molecular fraction
is set by the mid-plane pressure, we then expect lower $R_{\rm mol}$,
$\Sigma_{\rm SFR}$, and SFE in the low-$v_{\rm circ}$ galaxy.

In this paper, we address whether there is a SFE transition at $v_{\rm
circ} = 120 \, {\rm km \, s^{-1}}$ in our sample of bulgeless disk
galaxies (Section~\ref{sec:no_trans}).  We also investigate whether
scale height differences affect the SFE and discuss implications for
the scale of physical processes that are important for setting the
molecular fraction and the SFE.  We examine our results in light of
recent star formation models such as the the mid-plane pressure model
and the model of \citet{krumholz09} (Sections~\ref{sec:stability} and
\ref{sec:metallicity}).  To carry out this study, we trace molecular
gas with CO(1--0) data from the Institut de Radioastronomie
Millim\'etrique (IRAM) $30 \, {\rm m}$ telescope and atomic gas with
\ion{H}{1} $21 \, {\rm cm}$ data from the Very Large Array (VLA;
Watson et al. 2011, hereafter Paper~I).  We trace the SFR with
H$\alpha$ data from the 2.4~m Hiltner Telescope of the MDM Observatory
combined with polycyclic aromatic hydrocarbon (PAH) emission data
derived from {\it Spitzer Space Telescope} Infrared Array Camera
(IRAC) observations.  We also estimate the stellar mass from the {\it
Spitzer} IRAC data.  These observations, and measurements derived from
the data, are described in Section~\ref{sec:sample}.  We describe the
quantities that we derive from these measurements in
Section~\ref{sec:derived_quantities}.  Our results are in
Section~\ref{sec:results} and we discuss these results in
Section~\ref{sec:discussion}.

\section{Sample Summary, Observations, Data Reductions, and Measurements}
\label{sec:sample}
Our sample is composed of 20 Sd-Sdm galaxies within $32 \, {\rm Mpc}$,
with circular velocities between 46 and $190 \, {\rm km \, s^{-1}}$.
These properties are well matched to the \citet{dalcanton04} sample.
However, in contrast to the \citet{dalcanton04} sample of edge-on
galaxies, we selected our galaxies to be moderately inclined, with
inclinations between $16 \degr$ and $56 \degr$, such that we can
accurately measure the SFR and gas surface densities and place the
galaxies on the star formation law.  Section~2 in Paper~I and
Table~\ref{tab:sample} provide a description of the sample selection.

Many of the measurements described in this section were carried out to
derive surface densities -- of gas, SFR, and stars.  These surface
densities must be measured over the same area.  The IRAM CO(1--0) data
are the limiting factor, as they are single-beam, ${\rm FWHM} =
21\arcsec$ measurements centered on each galaxy.  Therefore, we
measured the emission within a 21$\arcsec$-diameter circular aperture
centered on the IRAM pointing center, the coordinates of which are
listed in Table~\ref{tab:obs_prop}, for the following datasets: the
\ion{H}{1} data from the VLA, the H$\alpha$ data from the MDM
Observatory, and the PAH and $4.5 \, \mu {\rm m}$ data from {\it
Spitzer} IRAC.

\subsection{IRAM 30 m CO(1--0)}
\label{sec:CO}
Thirteen of our objects were observed in the CO(1--0) and CO(2--1)
lines at 115 and 230\,GHz in May 2007 with the IRAM $30 \, {\rm m}$
telescope on Pico Veleta.  Dual polarization receivers were used at
both frequencies with the 512 $\times$ 1 MHz filterbanks on the
CO(1--0) line and the 256 $\times$ 4 MHz filterbanks on the CO(2--1)
line.  The observations were carried out in wobbler switching mode
with a wobbler throw of 200\arcsec \ in the azimuthal direction.  At
the beginning of the observations the frequency tuning was checked by
observing a bright galaxy at a similar redshift. Observations of the
same calibration source on different days allowed us to check the
relative calibration, which was excellent (better than 10\%) for
CO(1--0).  The calibration in CO(2--1) was equally good, except for
one day when the calibration observation was different by $\sim 35\%$.
Pointing was monitored on nearby quasars, Mars, or Jupiter every 60 --
90 minutes.  During the observation period, the weather conditions
were generally good, with pointing better than $3\arcsec$. The typical
system temperature was 300-500 K at 115\,GHz and 500--1000\,K at
230\,GHz on the $T_{\rm A}^*$ scale.  At 115 GHz (230 GHz), the IRAM
forward efficiency, $F_{\rm eff}$, was 0.95 (0.91), the beam
efficiency, $B_{\rm eff}$, was 0.75 (0.54), and the half-power beam
size is 21$\arcsec$ (11$\arcsec$).  All CO spectra and line
intensities are presented on the main beam temperature scale ($T_{\rm
mb}$) which is defined as $T_{\rm mb} = (F_{\rm eff}/B_{\rm
eff})\times T_{\rm A}^*$.  For the data reduction, we selected the
observations with good quality (taken during satisfactory weather
conditions and showing a flat baseline), averaged the spectra from the
individual scans of the source, and subtracted a constant continuum
for the CO(1--0) spectra and a linear continuum for the CO(2--1)
spectra.

Figure~\ref{fig:IRAM_spec_1} shows the CO spectra for the objects
observed in May 2007.  The black spectrum shows the CO(1--0) data and
the red spectrum shows the CO(2-1) data.  We did not use the CO(2-1)
data in this paper, but show it for completeness and to corroborate
some of the weaker CO(1--0) detections.  The solid black horizontal
line is the velocity range over which we integrated to derive the
CO(1--0) line intensity. The dashed horizontal line is centered on the
systemic velocity of the galaxy, derived from velocity field modeling
of our \ion{H}{1} emission line data.  The width of the dashed line is
the width of the \ion{H}{1} emission line at 20\% of the peak flux
density ($W_{20}$; see Table~\ref{tab:sample} for these values).  The
integrated CO(1--0) line intensities are presented in
Table~\ref{tab:obs_prop}.

We did not detect ESO~544-G03, ESO~418-G008, ESO~501-G023, or
UGC~6446.  For these galaxies, we quoted the CO line intensities as
upper limits, computed with $I_{\rm CO} < 3 \, \sigma_{\rm rms} \,
(W_{20} \, \Delta v)^{1/2}$, where $\sigma_{\rm rms}$ is the noise in
the spectrum in K and $\Delta v$ is the CO(1--0) spectrum channel
width, which is $10 \, {\rm km \, s^{-1}}$ in the Hanning-smoothed
spectra of Figure~\ref{fig:IRAM_spec_1}.

CO(1--0) and CO(2--1) spectra and line intensities for six of our
objects -- PGC~3853, NGC~2805, NGC~3906, NGC~4519, NGC~5964, and
NGC~6509 -- were presented in \citet{boeker03}, also based on IRAM $30
\, {\rm m}$ observations and reduced in the same manner.  We did not
obtain CO data for ESO~555-G027.

\begin{figure*}
\begin{center}
\plotone{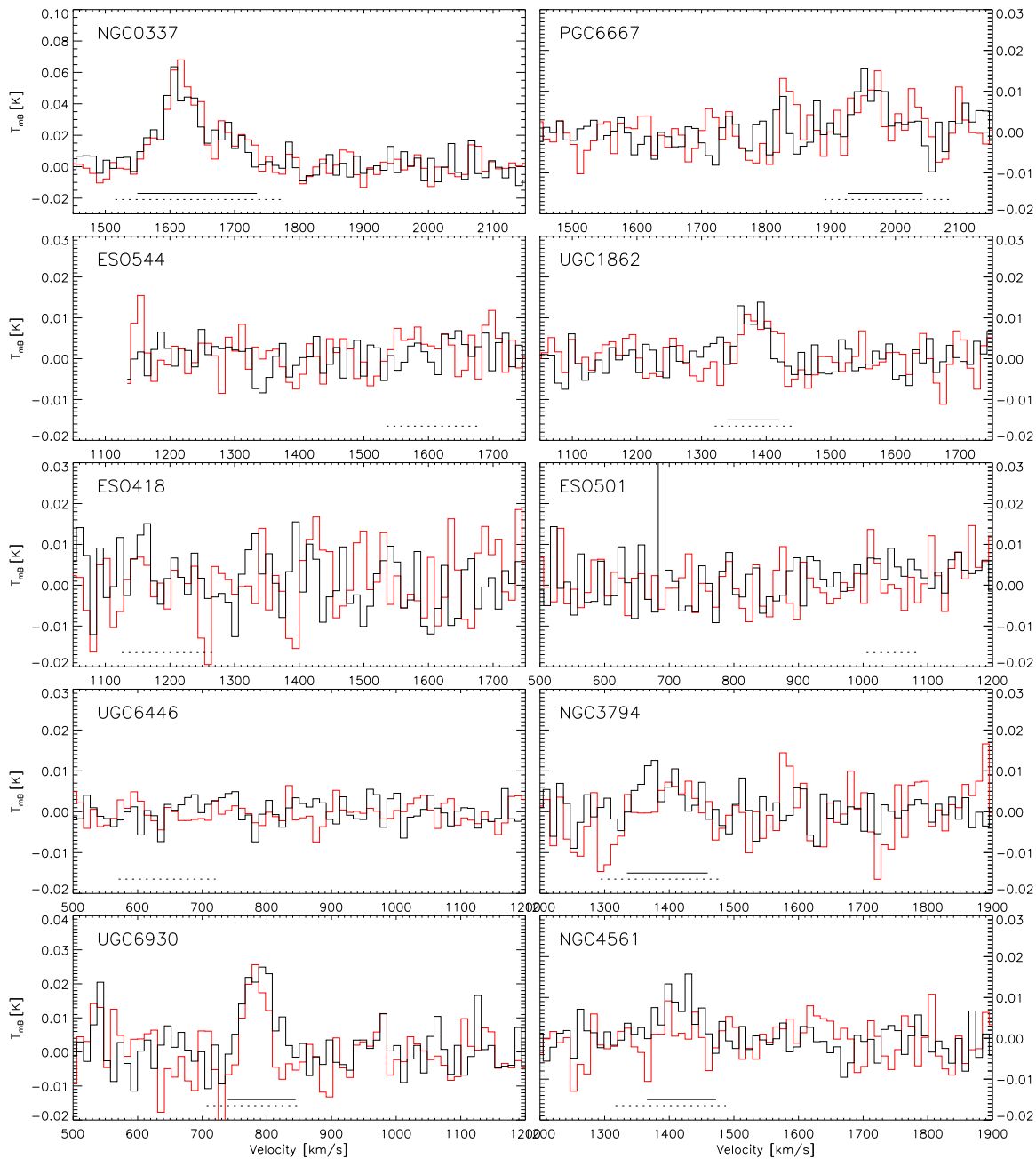}
\caption{CO(1--0) (black) and CO(2--1) (red) spectra for the objects
  observed in May 2007.  See \citet{boeker03} for the remaining
  spectra.  The black horizontal line designates the velocity range
  over which we integrated to derive the CO(1--0) line intensity.
  ESO~544-G03, ESO~418-G008, ESO~501-G023, and UGC~6446 do not have
  this line because they were undetected.  The dashed horizontal line
  is centered on the systemic velocity derived from velocity field
  modeling of our \ion{H}{1} data, and has a width of $W_{20}$, which
  we derived from the integrated \ion{H}{1} line profile.}
\label{fig:IRAM_spec_1}
\end{center}
\end{figure*}

\subsection{VLA HI}
\subsubsection{Integrated \ion{H}{1} Line Intensity}
\label{sec:HI}
The \ion{H}{1} $21 \, {\rm cm}$ data for our sample were obtained from
the VLA/Expanded VLA, operated by the National Radio Astronomy
Observatory\footnote{The National Radio Astronomy Observatory is a
facility of the National Science Foundation operated under cooperative
agreement by Associated Universities, Inc.}, for projects AZ0133
(carried out in 2001 August), AL0575 (carried out in 2002 June and
November), AM0873 (carried out in 2006 October and November), and
AM0942 (carried out in 2008 May and 2009 July and August).  The
galaxies were observed in the C or CnB configurations, which provide a
nominal angular resolution of $13\arcsec$.  The channel width is
generally $5.2 \, {\rm km \, s^{-1}}$.  The observations and
reductions are described further in Paper~I.  We measured the
integrated \ion{H}{1} line intensity from velocity-integrated
intensity maps (with units of ${\rm Jy \, beam^{-1} \, km \, s^{-1}}$)
created from naturally-weighted data cubes.  The beam major axis FWHM
(${B_{\rm maj}}$) and minor axis FWHM (${B_{\rm min}}$) for each data
cube are listed in Table~3 of Paper~I.

\begin{figure*}
\figurenum{1}
\begin{center}
\plotone{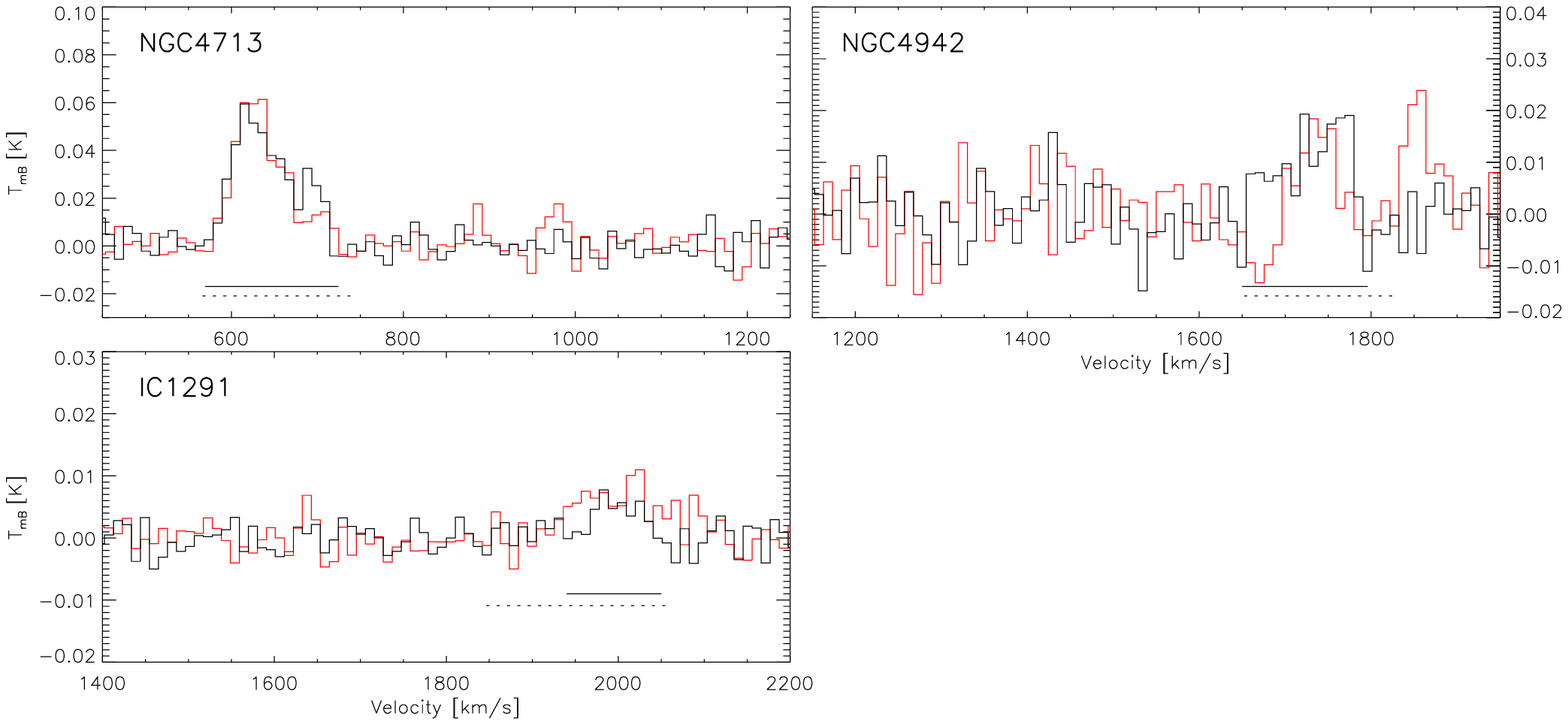}
\caption{(Continued)}
\label{fig:IRAM_spec_1}
\end{center}
\end{figure*}

Ten of our objects have $B_{\rm maj}$ and $B_{\rm min} < 21 \arcsec$.
To adjust the \ion{H}{1} beam to match the CO(1--0) beam, we used the
AIPS task CONVL to convolve the integrated intensity map so the output
Gaussian beam has ${B_{\rm maj}} = {B_{\rm min}} = 21 \arcsec$.  To
approximately match the \ion{H}{1} beam to the CO(1--0) beam for the
remaining ten objects with $B_{\rm maj} > 21 \arcsec$ (on average,
$B_{\rm maj} = 29\arcsec$) and $B_{\rm min} < 21 \arcsec$, we either
used the original integrated intensity map or convolved the map to
have $B_{\rm min} = 21 \arcsec$ and the beam major axis approximately
equal to the original $B_{\rm maj}$.

Using the resulting maps, we measured the average \ion{H}{1} line
intensity in ${\rm Jy \, beam^{-1} \, km \, s^{-1}}$ within the $21
\arcsec$-diameter circular aperture ($\langle {I_{\rm HI}} \rangle$)
with the AIPS task IMEAN.  These values are listed in
Table~\ref{tab:obs_prop}.  The integrated \ion{H}{1} line intensity in
${\rm K \, km \, s^{-1}}$ is given by:
\begin{equation}
\langle {I_{\rm HI}} \rangle \, [{\rm K \, km \, s^{-1}}] = 
\frac{6.07 \times 10^{5} \ \langle I_{\rm HI} \rangle\, [{\rm Jy \,
beam^{-1} \, km \, s^{-1}}]}{B_{\rm maj} \, B_{\rm min}},
\end{equation}
where ${B_{\rm maj}}$ and ${B_{\rm min}}$ are in arcseconds and now
refer to the beam of the map on which we made the measurements.  These
values are listed in Table~\ref{tab:obs_prop}.

For the objects with ${\rm B_{maj}} > 21 \arcsec$, we assumed that the
average line intensity within a $21 \arcsec$ beam is equal to the
measured line intensity from our image with a larger beam.  We
estimated the uncertainty introduced by this assumption by convolving
the integrated intensity maps of three objects with ${\rm B_{maj}} <
21 \arcsec$ such that the convolved beams match those of the ten
objects with ${\rm B_{maj}} > 21 \arcsec$.  We measured the average
line intensity within the $21 \arcsec$-diameter circular aperture and
compared this to the true value measured from the map with a $21
\arcsec$ beam.  We found that the line intensities in our test cases
differ from the true values by up to 11\% (in the case of the NGC~4519
beam, which has a $B_{\rm maj} = 51.91\arcsec$) and by 4\% on average.
The test measurements both over and under estimate the true value
depending on the emission distribution.  Therefore, we included this
in our uncertainty estimate, but make no correction.

The main contributors to the final uncertainty in our \ion{H}{1} line
intensities are flux calibration (5\%), aliasing (up to 11\% and
described in Paper~I), and using an image with a beam larger than
$21\arcsec$ to estimate the line intensity within $21 \arcsec$ (on
average 4\%).  Not all objects are subject to the latter two
uncertainties.  Nonetheless, we conservatively assigned the quadrature
sum of these uncertainties (13\%) as the generic uncertainty
associated with our \ion{H}{1} line intensities within the $21
\arcsec$-diameter aperture.

NGC~6509 shows \ion{H}{1} in absorption on the east side of the galaxy
because it is in the foreground of the radio source 4C~+06.63.  The
average line intensity within $21\arcsec$ is unaffected because the
eastern edge of the aperture and the western edge of the radio lobe
are separated by about $20\arcsec$.

\newpage

\subsubsection{Epicyclic Frequency}
\label{sec:ep_freq}
We calculated a representative epicyclic frequency ($\kappa$) for the
$21\arcsec$-diameter circular aperture to use in the stability
analysis of Section~\ref{sec:stability}.  We used $\kappa =
\sqrt{2(1+\beta)} \, (v/r)$, where $\beta = d{\rm log}(v)/d{\rm
log}(r)$.  We determined $\beta$ and the rotation velocity, $v$, at a
radius, $r$, of $5.25\arcsec$ (half the radius of the
$21\arcsec$-diameter aperture) from the \ion{H}{1} rotation curves
presented in Paper~I, where we fit a tilted ring model to the data to
derive $v$ at radii every $(B_{\rm maj} B_{\rm min})^{0.5}$, beginning
at $(B_{\rm maj} B_{\rm min})^{0.5}/2$.  We simply fit a line between
the origin and the first point in the rotation curve, the average
radius of which is 8.3$\arcsec$.  We used the slope of the line as an
estimate of $dv/dr$ and evaluated $v$ at $5.25\arcsec$.  There may be
inaccuracies introduced to $\kappa$ evaluated in this manner because
the origin and first point in the rotation curve are within a single
beam.  Therefore, we also calculated the epicyclic frequency at
$10.5\arcsec$ using the same method as above, but by fitting the line
between the first and second points from the rotation curve, where the
average second radius is $25.0\arcsec$.  The epicyclic frequencies
calculated at $10.5\arcsec$ are smaller than the values at
$5.25\arcsec$ by 30\% on average.  We used the epicyclic frequencies
evaluated at $5.25\arcsec$ in our stability analysis (and list these
in Table~\ref{tab:der_prop}), but include 30\% uncertainties on the
values.  Beam smearing, where many velocity components are within the
spatial beam, is likely in effect in this region.  Beam smearing leads
to underestimated velocities and gradients \citep{swaters99} and
therefore $\kappa$ may also be underestimated.  To account for this,
we use $\kappa$ evaluated at $5.25\arcsec$, as these values are
larger.

\subsection{MDM \halpha}
H$\alpha$ and continuum images of the galaxies were obtained at the
MDM 2.4m Hiltner telescope over the course of four observing runs in
January 2007, November 2007, May/June 2007, and January 2008. Each
galaxy was observed for between 30 min and 2.5 hours through a pair of
matched, custom-made $15 \, {\rm nm}$ wide narrowband filters centered
at $663 \, {\rm nm}$ and $693 \, {\rm nm}$, hereafter the 663bp15 and
693bp15 filters, respectively. The H$\alpha$ emission line falls
within the 663bp15 bandpass for all of the galaxies in this
sample. Observations were obtained with the Direct CCD camera
``Echelle,'' which has 2048x2048 pixels. The CCD was binned over 2x2
pixels to produce a plate scale of $0.55"$/pixel, which was
well-matched to the $1''-1.5''$ image quality measured on most
nights. The field of view was $9.4'$. Conditions were photometric for
most of the observations and a series of spectrophotometric standards
were observed for flux calibration, as were a series of twilight
flats. The exposure times and observation dates are listed in
Table~\ref{tbl:halphalog}.

Overscan subtraction, flat fielding, cosmic ray rejection, and bad
pixel removal were performed with IRAF\footnote{IRAF is distributed by
the National Optical Astronomy Observatory, which is operated by the
Association of Universities for Research in Astronomy (AURA) under
cooperative agreement with the National Science Foundation.}. All of
the 663bp15 and 693bp15 images of each galaxy were registered to a
common coordinate system with SCAMP \citep{bertin06} and a combined
image for each filter was constructed with SWARP \citep{bertin02}.

Observations of spectrophotometric standard stars were used to
determine the absolute flux calibration for the 663bp15 filter
(including an airmass correction) and the relative throughput of the
two filters. We calculated the expected ratio of stellar flux in these
filters for late-type galaxies by using the SYNPHOT package in IRAF to
convolve the filter transmission functions with a series of Bruzual \&
Charlot 1995 \citep[e.g.,][]{bruzual03} stellar population synthesis
models with both continuous and exponentially declining ($\tau = 1$
Gyr) star formation histories. The range of flux ratios from these
models and the relative throughput of the two filters were used to
scale the 693bp15 image and subtract it from the 663bp15 image to
create an H$\alpha$ image for each galaxy.

We measured the H$\alpha$ flux for each galaxy within the
$21\arcsec$-diameter circular aperture and within a circle of diameter
$D_{25}$, the B-band major isophotal diameter at $25 \, {\rm mag \,
arcsec^{-2}}$, with aperture photometry. All H$\alpha$ flux
measurements were multiplied by a factor of 0.75 to account for
emission from [\ion{N}{2}] $\lambda\lambda 6548,6584$ in the 663np15
bandpass \citep{kennicutt83}. The H$\alpha$ flux measurements were
also corrected for Galactic extinction using the extinction law of
\citet{odonnell94} assuming $R_{V} = 3.1$ and reddening values from
\citet{schlegel98} and tabulated on NED\footnote{The NASA/IPAC
Extragalactic Database (NED) is operated by the Jet Propulsion
Laboratory, California Institute of Technology, under contract with
the National Aeronautics and Space Administration.}. The final values
within the $21\arcsec$ aperture are listed in
Table~\ref{tab:obs_prop}.

A number of our galaxies show substantial variation in the H$\alpha$
flux within the $21\arcsec$ aperture if we vary the center of the
aperture by the $\sim 2 - 3\arcsec$ pointing accuracy of the IRAM $30
\, {\rm m}$.  Including this uncertainty, as well as uncertainty in
the [NII] correction and absolute calibration uncertainties, we
estimate the uncertainty in the H$\alpha$ flux within the
$21\arcsec$-diameter aperture to be about 30\%.

\begin{figure}
\epsscale{1.2}
\begin{center}
\plotone{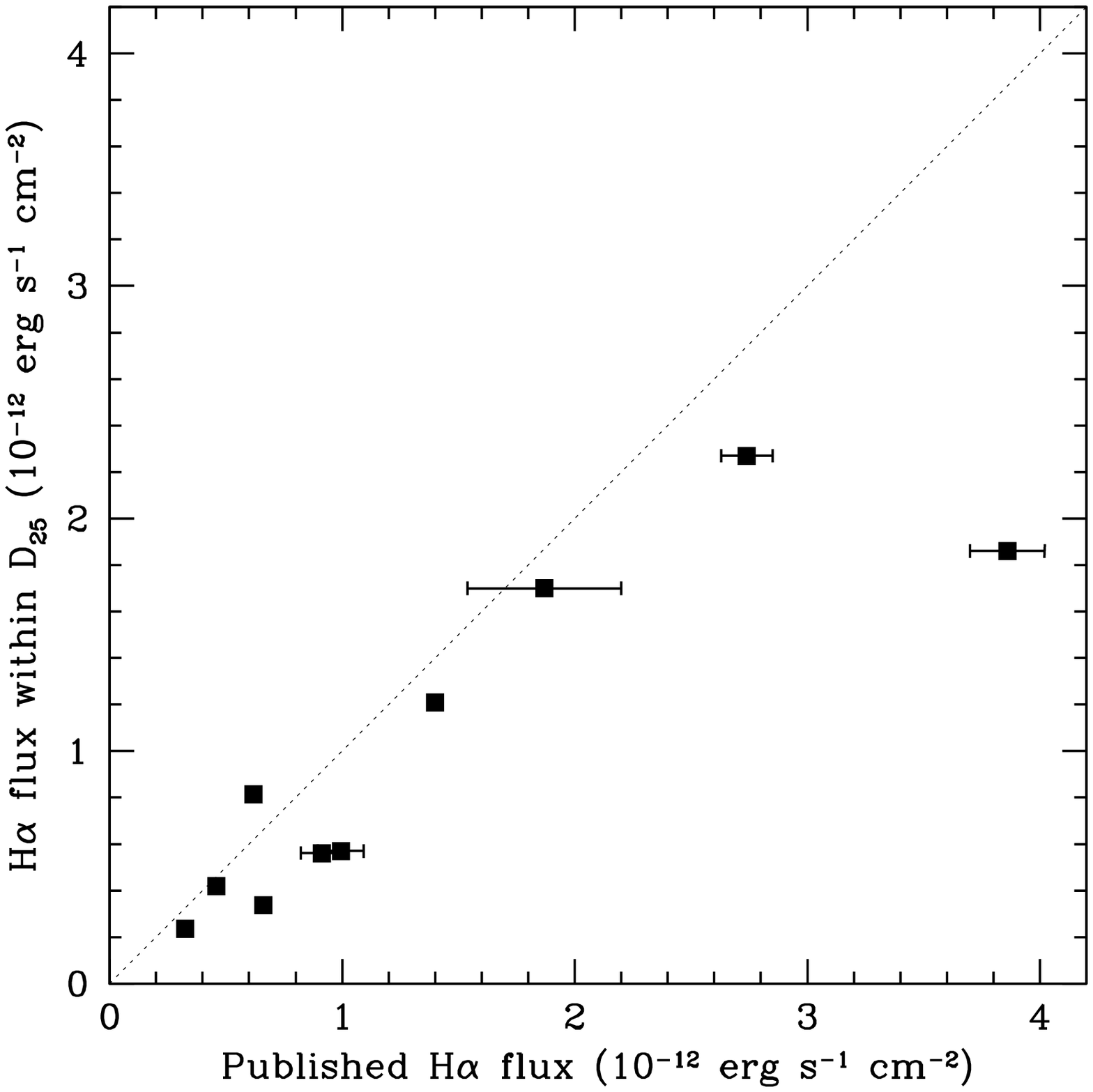}
\caption{Comparison between our H$\alpha$ fluxes measured within a
  circle of diameter $D_{25}$ and published H$\alpha$ fluxes.  Both
  values are corrected for [\ion{N}{2}] emission and Galactic
  extinction.  The dotted line shows equality.  The largest outlier is
  NGC~0337.}
\label{fig:Halpha_comp}
\end{center}
\end{figure}

Figure~\ref{fig:Halpha_comp} compares our H$\alpha$ fluxes measured
within $D_{25}$ relative to published values from \citet{koopmann01},
\citet{james04}, \citet{moustakas06}, and \citet{epinat08}, where we
corrected the published values for [\ion{N}{2}] emission and Galactic
extinction if necessary.  Our values are on average 20\% smaller than
the published values.  This systematic offset is comparable to our
calibration uncertainty and may be due to differences in the filter
profiles, the angular extent of the apertures, and the treatment of
bright stars with potentially large subtraction residuals. We note
that this offset corresponds to an uncertainty in the star formation
rate surface density of less than $0.1 \, {\rm dex}$, which is smaller
than the $0.2 \, {\rm dex}$ uncertainty that we assign due to the
variable contribution to dust heating from non-star forming
populations.

\subsection{{\it Spitzer} IRAC}
Fourteen of the 20 galaxies were observed as part of our {\it Spitzer}
Cycle 5 Program 50102; the remaining six (NGC~337, UGC~1862,
ESO~418-G008, NGC~2805, NGC~4519, and NGC~4561) were observed for
various other programs. Observations for Program 50102 consisted of
five, dithered observations with a frame time of 30s in all four IRAC
channels ($3.6 \, \mu {\rm m}$, $4.5 \, \mu {\rm m}$, $5.8 \, \mu {\rm
m}$, and $8 \, \mu {\rm m}$). Data for the other six archival datasets
were comparable to our observations, with the exception that the data
for UGC~1862 only include the galaxy in channels 2 and 4. In the
analysis described below, we use the channel 1, 2 and 4 data.

The Basic Calibrated Data (BCD) for all 20 galaxies were processed
with Sean Carey's artifact mitigation
software\footnote{http://spider.ipac.caltech.edu/staff/carey/irac\_artifacts/},
which corrects for a variety of effects such as muxbleed, column
pulldown/pullup, electronic banding, and first frame effect. We then
created mosaics for each channel with the {\it Spitzer Science
Center's} MOPEX (MOasicker and Point source EXtractor) package. This
package corrects individual images for background variations and
optical distortions, and then projects them onto an output mosaic
image for each channel.  These mosaic images were used for five
measurements for each galaxy: the inclination ($i$), position angle of
the major axis (PA), the PAH flux density, the $4.5 \, \mu {\rm m}$
flux density, and the exponential disk scale length.

\subsubsection{Inclination and Position Angle}
\label{sec:ellipse}
We estimated the inclination and PA for each galaxy with the IRAF {\tt
ellipse} task, which fits elliptical isophotes with the iterative
method described by \citet{jedrzejewski87}. The $3.6 \, \mu {\rm m}$
data ($4.5 \, \mu {\rm m}$ for UGC~1862) were fit because this channel
has the greatest sensitivity to the old stellar population, yet is
relatively insensitive to dust.

We found that the isophote fits, particularly in the outer,
low-surface brightness regions, were relatively sensitive to the
presence of bright stars.  We therefore created masks of these stars
with the SExtractor package \citep{bertin96} and included this mask as
an input to the {\tt ellipse} task. These masks also excluded regions
with relatively poor coverage in the IRAC mosaic. We averaged the PA
and ellipticity values for the largest isophotes to derive our final
PAs and inclinations. These values were used as inputs for the
rotation curve analysis described in Paper~I and were reported in
Table~5 of that work.

\subsubsection{PAH Flux Density}
We calculated the PAH flux density within the $21\arcsec$-diameter
circular aperture using the $8 \, \mu {\rm m}$ images, which are
dominated by the 7.7 and $8.6 \, \mu {\rm m}$ PAH features, and the
stellar emission-dominated $3.6 \, \mu {\rm m}$ images.  We measured
the $8 \, \mu {\rm m}$ and $3.6 \, \mu {\rm m}$ flux densities using
the IRAF task {\tt phot}.  In both measurements, we subtracted from
each pixel the median sky background, which we measured within a
large-radius annulus centered on the galaxy.  We applied the
band-specific extended source aperture correction to the $8 \, \mu
{\rm m}$ and $3.6 \, \mu {\rm m}$ flux densities (described in the
IRAC Instrument Handbook; for the $21\arcsec$-diameter circular
aperture, the correction is 0.985 and 0.896 for the $3.6 \, \mu {\rm
m}$ and $8 \, \mu {\rm m}$ bands, respectively).  Finally, the PAH
flux density is the $8 \, \mu {\rm m}$ flux density, less the stellar
emission contribution, which we estimated by scaling the $3.6 \, \mu
{\rm m}$ flux density by 0.255 \citep[as used in][]{kennicutt09}.  For
UGC~1862, we used the $4.5 \, \mu {\rm m}$ flux density, scaled by a
factor of 0.389, to remove the stellar contribution to the $8 \, \mu
{\rm m}$ band.  This scale factor is derived from values quoted in
\citet{helou04}, except we assumed that all the $4.5 \, \mu {\rm
m}$-band emission is stellar.  The uncertainties on the calibrated
$3.6 \, \mu {\rm m}$ and $8 \, \mu {\rm m}$ flux densities are about
10\% \citep{hora04}.  We quote PAH flux density uncertainties that are
simple error propagated values.  Our PAH flux densities are presented
in Table~\ref{tab:obs_prop}.

\subsubsection{$4.5 \, \mu {\rm m}$ Flux Density}

We calculated the total $4.5 \, \mu {\rm m}$ flux density and the flux
density within the $21\arcsec$-diameter circular aperture such that we
can derive the total stellar mass and the stellar mass surface density
in Section~\ref{sec:mass}.  We first masked out bright foreground
stars with the IRAF APPHOT and DAOPHOT packages, in particular the
{\tt phot} and {\tt substar} tasks.  We used large $4.5 \, \mu {\rm
m}$ to $8.0 \, \mu {\rm m}$ flux ratios to identify foreground stars.
For a few bright stars that were not adequately subtracted, we
manually replaced the affected pixels with values from a region at a
similar radius from the center of the galaxy.

We used the IRAF task {\tt phot} to measure the flux density within
the $21\arcsec$-diameter circular aperture.  For the total flux
density, we used the IRAF task {\tt ellipse} to measure the flux
density within an ellipse where the semi-major axis (SMA) is
$D_{25}/2$ and the PA and ellipticity ($e = 1-{\rm cos} \, i$) were
derived from a combination of rotation curve analyses on the VLA
\ion{H}{1} data and the ellipse fits of the IRAC $3.6 \, \mu {\rm m}$
data, described in Section~\ref{sec:ellipse} (PA and $i$ are given in
Table~\ref{tab:sample}).  In both measurements, we subtracted the
median sky background from each pixel, measured within a large-radius
annulus centered on the galaxy.

To confirm that the $D_{25}$ aperture is appropriate for the total
flux density measurement, we identified the SMA where the surface
brightness profile flattens (specifically, where the surface
brightness decreases by, on average, less than 2\% over a number of
apertures).  The flux density enclosed within this ellipse was
generally within 7\% of the flux density within the $D_{25}$ ellipse,
except in two cases: UGC~6446 and UGC~6930, where the flux density
within the $D_{25}$ ellipse was larger by a factor of 1.9 and 1.5,
respectively.  We henceforth use the flux density within the $D_{25}$
ellipse as the total flux density.

We applied the extended source aperture correction to the $21 \arcsec$
and $D_{25}$ flux density measurements, with a value of 1.013 for the
$21\arcsec$ aperture and an average value of 0.944 for the $D_{25}$
measurements.  We did not apply a color correction.  The $21 \arcsec$
and total $4.5 \, \mu {\rm m}$ flux densities are listed in
Table~\ref{tab:obs_prop}.

\subsubsection{Exponential Disk Scale Length}
\label{sec:scale_lengths}
We estimated the exponential scale length of each galaxy using the
IRAF {\tt ellipse} task on the $4.5 \, \mu {\rm m}$ data, allowing the
center, PA, and ellipticity to vary as a function of semi-major axis.
We fit an exponential to the mean isophotal intensity profile to
derive the central surface brightness and scale length.  We excluded
PSF and/or bar components from the profile fit based on visual
examination of the images and a provisional GALFIT \citep{peng02}
analysis.  We created a two-dimensional image representing the {\tt
ellipse} profile with {\tt bmodel}, subtracted the model from the
original image, and found that the standard deviation in the region of
the galaxy within the residual image is typically less than about ten
times the standard deviation in a galaxy-free region of the original
image.  Given the small-scale structures present in most of the
galaxies, we accepted these values and fits.  The scale lengths are
listed in Table~\ref{tab:der_prop} and are between 0.69 and $3.4 \,
{\rm kpc}$.

We assigned an error of 20\% to the scale lengths, based on comparing
the scale lengths computed as described above to scale lengths
computed from ellipse runs where we held the PA and ellipticity fixed
as a function of semi-major axis at the values from
Table~\ref{tab:sample}.  We confirmed that scale lengths derived from
the $3.6 \, \mu {\rm m}$ data are generally consistent with the values
derived from the $4.5 \, \mu {\rm m}$ data within the uncertainty
(they differ by 9\% on average).

\section{Derived Quantities}
\label{sec:derived_quantities}
For all the surface density calculations below, we used the
$21\arcsec$-diameter aperture, to match the beam of our CO(1--0) data
from IRAM.  Table~\ref{tab:der_prop} lists the physical size of
$21\arcsec$ (0.7 - $3.2 \, {\rm kpc}$) and the parameters derived in
the following sections.  The surface densities are all within the
deprojected area of the aperture, which we calculated with the
inclinations from Table~\ref{tab:sample}.

\subsection{Atomic, Molecular, and Total Hydrogen Surface Density}
\label{sec:gas_SD}
The \ion{H}{1} surface density is given by:
\begin{equation}
\Sigma_{\rm HI} \, [M_{\odot} \, {\rm pc^{-2}}] = 0.015 \, \langle
I_{\rm HI} \rangle \, {\rm cos}(i),
\end{equation}
where $\langle I_{\rm HI} \rangle$ is the average integrated
\ion{H}{1} line intensity within the $21\arcsec$-diameter circular
aperture in ${\rm K \, km \, s^{-1}}$ from Section~\ref{sec:HI}.  We
did not include a correction for He.  The $\Sigma_{\rm HI}$
uncertainty is dominated by the contribution from the line intensity
uncertainty and the typical uncertainty in ${\rm log} \, (\Sigma_{\rm
HI})$ is $0.06 \, {\rm dex}$.  The ${\rm H_{2}}$ surface density is
given by:
\begin{multline}
\Sigma_{\rm H_{2}} \, [M_{\odot} \, {\rm pc^{-2}}] = \\
3.2 \, \frac{X_{\rm CO}}{2.0 \times 10^{20} \, {\rm cm^{-2} \, (K \,
  km \, s^{-1})^{-1}}} \, I_{\rm CO} \, {\rm cos}(i),
\end{multline}
where $X_{\rm CO}$ is the CO-to-${\rm H}_{2}$ conversion factor and
$I_{\rm CO}$ is the CO line intensity in ${\rm K \, km \, s^{-1}}$
from Section~\ref{sec:CO}. We used a constant $X_{\rm CO}$ of $2.8
\times 10^{20} \, {\rm cm^{-2} \, (K \, km \, s^{-1})}^{-1}$.  We used
this value rather than the Milky Way value of $2.0 \times 10^{20} \,
{\rm cm^{-2} \, (K \, km \, s^{-1})}^{-1}$ so we can plot the
\citet{kennicutt98} total hydrogen star formation law relative to our
data.  Again, we did not include a correction for He.

The typical uncertainty in ${\rm log} \, (\Sigma_{\rm H_{2}})$ is
$0.07 \, {\rm dex}$, due mainly to the CO line intensity uncertainty.
We did not include uncertainty due to $X_{\rm CO}$.  \citet{leroy11}
studied five local group galaxies and concluded that $X_{\rm CO}$ is
relatively constant at the solar value for $12+{\rm log} \, (O/H)
\gtrsim 8.4$ and increases with decreasing oxygen abundance below
$12+{\rm log} \, (O/H) \sim 8.2 - 8.4$.  We have only one galaxy where
we estimated that the oxygen abundance is below $12+{\rm log} \, (O/H)
= 8.4$ (Section~\ref{sec:mass}), so we do not expect much $X_{\rm CO}$
variation in our sample.

We also use the total hydrogen surface density, $\Sigma_{\rm HI+H_{2}}
= \Sigma_{\rm HI} + \Sigma_{\rm H_{2}}$.  The typical uncertainty in
${\rm log} \, (\Sigma_{\rm HI+H_{2}})$ is $0.05 \, {\rm dex}$.

\subsection{Star Formation Rate Surface Density}
\label{sec:SFR_SD}
We used H$\alpha$ emission to trace the unobscured SFR and PAH
emission to trace the obscured SFR.  The H$\alpha$ emission is due to
recombination in \ion{H}{2} regions, which are ionized by O and early
B stars.  PAHs are small dust grains (or large molecules) that are
primarily excited by single UV photons \citep{sellgren84, leger84}.
Because ionizing radiation is not required to heat PAHs \citep[and in
fact there is evidence that ionizing radiation destroys
PAHs;][]{helou04, peeters04}, PAH emission traces lower-mass,
longer-lived stars than those ultimately responsible for the H$\alpha$
emission. PAH emission has been used to trace the total SFR
\citep[e.g.,][]{zhu08}, but there is variation in the $8 \, \mu {\rm
m}$ luminosity at a given SFR due to environment, especially
metallicity \citep[e.g.,][]{calzetti07}.  \citet{kennicutt09} used 75
galaxies to calibrate SFR estimates based on H$\alpha$ and PAH
emission by comparing the combined H$\alpha$ and PAH luminosity to the
H$\alpha$ luminosity corrected for extinction using the Balmer
decrement.  The authors found that SFRs calculated with H$\alpha$ and
PAH emission agreed with their reference SFRs with as little scatter
as SFRs calculated with H$\alpha$ and $24 \, \mu {\rm m}$ emission.
We used the SFR calibration of \citet{kennicutt09} to calculate the
SFR surface density within the $21\arcsec$-diameter circular aperture:
\begin{multline}
\Sigma_{\rm SFR} [M_{\odot} \, {\rm yr^{-1} \, kpc^{-2}}]= \\
7.30 \times 10^{10}(F_{\rm H\alpha} + 1.1 \times 10^{-28} \, \nu \,
F_{\rm PAH}) \, {\rm cos}(i),
\end{multline}
where $F_{H\alpha}$ and $F_{\rm PAH}$ are the H$\alpha$ flux in ${\rm
erg \, s^{-1} \, cm^{-2}}$ and the PAH flux density in mJy within the
$21\arcsec$-diameter circular aperture, $\nu$ is the central frequency
of the $8\, \mu{\rm m}$ IRAC band in Hz, and the constant includes the
aperture area.  This assumes the initial mass function (IMF) from
\citet{calzetti07}, which is similar to that presented in
\citet{kroupa01}.  The \citet{kennicutt09} SFR was calibrated for
galaxy-averaged data, but should be appropriate for our data because
we generally probe regions that are a couple of square kpc in area and
should therefore contain a number of star forming regions.  Even using
both H$\alpha$ and PAHs to trace the SFR, galaxies with low
metallicity could have underestimated SFRs because the fraction of
dust mass in PAHs decreases at low metallicity, particularly below
$Z_{\odot}/4$ \citep[$12+{\rm log} \, (O/H) \lesssim
8.3$;][]{draine07,smith07}.  In the following section we estimate the
oxygen abundance of each galaxy from the stellar mass and find that
only one galaxy (ESO~501-G023) has $12+{\rm log} \, (O/H) < 8.3$.
Therefore, we do not expect significant SFR underestimates in our
sample.

\citet{kennicutt09} discussed that the dominant source of uncertainty
in their SFRs is due to varying contributions to dust heating from
older stellar populations ($\gtrsim 100 \, {\rm Myr}$).  Based on their
suggestions, and including the fact that our sample has a limited
range in morphology and therefore star formation history, we assigned
a general uncertainty of $0.2 \, {\rm dex}$ to our SFR surface
densities.  This dominates over the contribution due to H$\alpha$ flux
and PAH flux density measurement uncertainties.

In upcoming sections, we use the SFE defined as $\Sigma_{\rm
SFR}/\Sigma_{\rm HI+H_{2}}$ in ${\rm yr}^{-1}$.  The typical
uncertainty in the SFE is $0.2 \, {\rm dex}$.

\subsection{Stellar Mass Surface Density, Total Stellar Mass, and
  Oxygen Abundance}
\label{sec:mass}
We derived the stellar mass surface density within the $21
\arcsec$-diameter circular aperture and the total stellar mass from
the $4.5 \, \mu {\rm m}$ flux densities.  We chose to use the $4.5 \,
\mu {\rm m}$ data over the $3.6 \, \mu {\rm m}$ data because there are
no PAH emission features in the $4.5 \, \mu {\rm m}$ band.

To estimate the stellar mass, we used a relationship between K-band
mass-to-light ratio and color from \citet{bell01}, derived from
stellar population synthesis modeling:
\begin{equation}
{\rm log} \, (\Upsilon_{\ast}^{\rm K}) = 1.43(J-K_{\rm s})-1.53,
\label{eqn:one}
\end{equation}
where $\Upsilon_{\ast}^{\rm K}$ is the mass-to-light ratio in the K
band in $M_{\odot}/L_{\rm K,\odot}$, $L_{\rm K,\odot}$ is the solar
luminosity in the K band, and the $J$ and $K_{\rm s}$ magnitudes are
from the Two Micron All Sky Survey \citep[2MASS; mainly the Extended
Source Catalog, but three galaxies are in the Large Galaxy Atlas
of][]{jarrett03} and are listed in Table~\ref{tab:obs_prop}.  The
original relationship uses Johnson $K$ magnitudes, but use of $K_{\rm
s}$ magnitudes does not introduce significant error to the mass. This
relation is a linear combination of the $\Upsilon_{\ast}^{\rm K}$ -
$(V-J)$ and $\Upsilon_{\ast}^{\rm K}$ - $(V-K)$ relations presented in
Table~1 of \citet{bell01} and we subtracted $0.15 \, {\rm dex}$ to
convert from the scaled Salpeter IMF that the authors use to a
\citet{kroupa01} IMF.  For three galaxies that do not have 2MASS data,
we used:
\begin{equation}
{\rm log} \, \Upsilon_{\ast}^{\rm K} = 0.21(B-K_{\rm s})-1.11 =
0.21(B-[4.5]) -1.23,
\label{eqn:two}
\end{equation}
where we used B-band magnitudes from the Third Reference Catalogue of
Bright Galaxies \citep[RC3;][also listed in
Table~\ref{tab:sample}]{RC3}.  This equation is similar to that used
in \citet{lee06}, but we used the \citet{bell01} Table~1 model
results, we have converted to a \citet{kroupa01} IMF, and we assumed
$(K_{\rm s}-[4.5]) = 0.58$.  This color is the average value from the
eleven galaxies in our sample where the 2MASS aperture radius
(``r\_m\_ext'') and our aperture SMA ($D_{25}/2$) differ by less than
40\% of our SMA.  Six galaxies have spuriously large/red $(K_{\rm
s}-[4.5])$ colors because the 2MASS aperture radius is much smaller
than our aperture radius \citep[as small as 20\% of our aperture
radius; for a similar effect, see][]{dale09}.

We then calculated the stellar mass within the $21 \arcsec$-diameter
circular aperture and the total stellar mass with:
\begin{equation}
{\rm log} \, (M_{\ast}/M_{\odot}) = {\rm log} \, \Upsilon_{\ast}^{\rm
  K} + {\rm log} \, (L_{4.5}/L_{4.5,\odot}) - 0.4(K_{\rm s}-[4.5]),
\label{eqn:three}
\end{equation}
as in \citet{lee06}.  To calculate the luminosity of the galaxy,
$L_{4.5}$, we used either the 4.5$\, \mu {\rm m}$ flux density within
the $21\arcsec$-diameter circular aperture or the total 4.5$\, \mu
{\rm m}$ flux density from Table~\ref{tab:obs_prop}, the distances
presented in Table~\ref{tab:sample}, the absolute 4.5$ \, \mu {\rm m}$
magnitude of the Sun, $M_{4.5, \odot} = 3.3$ \citep{lee06, oh08}, and
the zero-magnitude flux density of the 4.5$\, \mu {\rm m}$ band,
$F_{4.5}^0 = 179.7 \, {\rm Jy}$ \citep{reach05}.  This equation uses
the fact that $(K-[4.5])\sim 0$ for the Sun.  We again assumed
$(K_{\rm s}-[4.5]) = 0.58$, for all galaxies in this case.  For the
stellar mass surface density ($\Sigma_{\ast}$), we then divided by the
deprojected area of the $21\arcsec$-diameter circular aperture.

The average offset between masses computed with
Equations~\ref{eqn:one} and \ref{eqn:two} is 0.2$\, {\rm dex}$, with
masses computed with Equation~\ref{eqn:two} generally smaller.  This
offset is smaller than our estimate of the uncertainty in the mass,
described further below.  The masses computed with
Equation~\ref{eqn:three} above are in good agreement with masses
computed using the \citet{oh08} relationship between K-band and 4.5$\,
\mu {\rm m}$ band mass-to-light ratios, derived from stellar
population synthesis modeling.

\citet{bell01} cite uncertainties of $0.1-0.2\, {\rm dex}$ in their
mass-to-light ratios, which includes model and dust uncertainties and
allows for small star formation bursts.  The uncertainty introduced to
the stellar mass by uncertainties in $L_{4.5}$ is of similar order,
mainly due to distance uncertainties.  We assigned a general
uncertainty of 0.3$\, {\rm dex}$ to the stellar masses, which is the
quadrature sum of the above uncertainties.  This is in agreement with
the \citet{conroy09} estimate of typical stellar mass uncertainties
from stellar population synthesis modeling.  The uncertainty in the
stellar mass surface density is dominated by the contribution from the
mass-to-light ratio.  We therefore assigned a typical uncertainty of
$0.2 \, {\rm dex}$.

We estimated a representative oxygen abundance for each galaxy to help
assess the accuracy and precision of our molecular hydrogen and SFR
surface densities (Sections~\ref{sec:gas_SD} and \ref{sec:SFR_SD}) and
also so we can compare our data to the \citet{krumholz09} model
(Section~\ref{sec:metallicity}). We estimated the oxygen abundance
with the total stellar masses calculated from the 4.5$\, \mu {\rm m}$
flux density and the mass-metallicity relation of \citet{tremonti04}
(their Equation~3), which assumes a \citet{kroupa01} IMF, consistent
with the rest of our analysis.  We derive an average uncertainty in
$12+{\rm log}(O/H)$ of $0.11 \, {\rm dex}$.

\subsection{Stability Parameters and Mid-Plane Pressure}
\label{sec:Q_calc}
In Section~\ref{sec:stability}, we study trends between SFE and
stability, parametrized by a generalized Toomre Q parameter
\citep{toomre64, rafikov01}.  We calculated the stability parameters
within the $21\arcsec$-diameter circular aperture.  The stability
parameter for the gas and stellar components of the disk are
\begin{equation}
Q_{\rm gas} = \frac{\kappa \sigma_{\rm gas}}{\pi {\rm G} \Sigma_{\rm
gas}}, {\rm and}
\end{equation}
\begin{equation}
Q_{\rm stars} = \frac{\kappa \sigma_{\ast, \rm r}}{\pi {\rm G}
\Sigma_{\ast}},
\end{equation}
respectively, where $\kappa$ is the epicyclic frequency and
$\sigma_{\rm gas}$ and $\sigma_{\ast, \rm r}$ are the gas and radial
stellar velocity dispersions.  \citet{rafikov01} calculated the
stability parameter that includes both gas and stars in a thin
rotating disk:
\begin{equation}
\label{eqn:Q}
\frac{1}{Q_{\rm gas+stars}} = \frac{2}{Q_{\rm stars}} \, \frac{q}{1+q^{2}} +
\frac{2}{Q_{\rm gas}} \, R \, \frac{q}{1+q^{2}R^{2}},
\end{equation}
where $q = k \sigma_{\ast, \rm r}/ \kappa$, $R = \sigma_{\rm
gas}/\sigma_{\ast, \rm r}$, and $k$ is a free parameter that
represents the wavenumber of the perturbation.  In all cases, the
instability condition is when $Q < 1$ (except, strictly speaking,
$Q_{\rm stars} < 1.07$).  \citet{romeo11} included disk thickness in
the stability estimation by incorporating a factor related to the
ratio of the vertical to radial velocity dispersion, but we settled on
the \citet{rafikov01} parameter for ease of comparison with other
works.

We generally made similar assumptions for the components of Q as
\citet{leroy08}.  We assumed $\sigma_{\rm gas} = 11 \, {\rm km \,
s^{-1}}$, which is appropriate for the warm neutral medium,
$\sigma_{\ast, \rm r} = 1.67 \sigma_{\ast, \rm z}$, and $\sigma_{\ast,
\rm z} = (2 \pi \, G \, l_{\ast} \, \Sigma_{\ast}/7.3)^{0.5}$, where
$\sigma_{\ast, \rm z}$ is the vertical stellar velocity dispersion and
$l_{\ast}$ is the stellar scale length
\citep{tamburro09,shapiro03,vanderkruit81,vanderkruit88}.  This
$\sigma_{\ast}$ equation assumes that the disk is isothermal in the z
direction, $h_{\ast}$ is constant as a function of radius, and
$l_{\ast}/h_{\ast} = 7.3$ \citep[as observed by][]{kregel02}, where
$h_{\ast}$ is the stellar scale height.  Note that the latter two
assumptions are not in conflict with a sample like \citet{dalcanton04}
because those authors found only a transition in dust scale height
with circular velocity, not in stellar scale height.  We used the
scale lengths and stellar mass surface densities from
Sections~\ref{sec:scale_lengths} and \ref{sec:mass} to derive
$\sigma_{\ast, {\rm r}}$ values between 9 and $78 \, {\rm km \,
s^{-1}}$.

For $\Sigma_{\rm gas}$, we multiplied the total-hydrogen surface
density ($\Sigma_{\rm HI+H_{2}}$) by a factor of 1.36 to include
helium.  For $\Sigma_{\ast}$, we used the value derived in
Section~\ref{sec:mass}.  For the epicyclic frequency, we used the
value from Section~\ref{sec:ep_freq}, evaluated at $5.25\arcsec$.  For
the wavenumber of the perturbation, $k = 2 \pi/\lambda$, where
$\lambda$ is the wavelength of the perturbation, we used the common
method of varying $\lambda$ to find the minimum $Q_{\rm gas+stars}$
($Q_{\rm gas+stars, min}$) for the region \citep[e.g.,][]{yang07,
yim11}.  This typically results in smaller, less stable $Q_{\rm
gas+stars}$ compared to $Q_{\rm gas}$ and $Q_{\rm stars}$.  We found
$\lambda = 0.5 - 3.8 \, {\rm kpc}$ at $Q_{\rm gas+stars, min}$.  We
propagated the uncertainties in $\kappa$, $l_{\ast}$, $\Sigma_{\rm
gas}$, and $\Sigma_{\ast}$ to derive a typical uncertainty in $Q_{\rm
gas}$ of 30\%, in $Q_{\rm stars}$ of 60\%, and in $Q_{\rm gas+stars}$
of 40\%.

\begin{figure}
\epsscale{1.2}
\begin{center}
\plotone{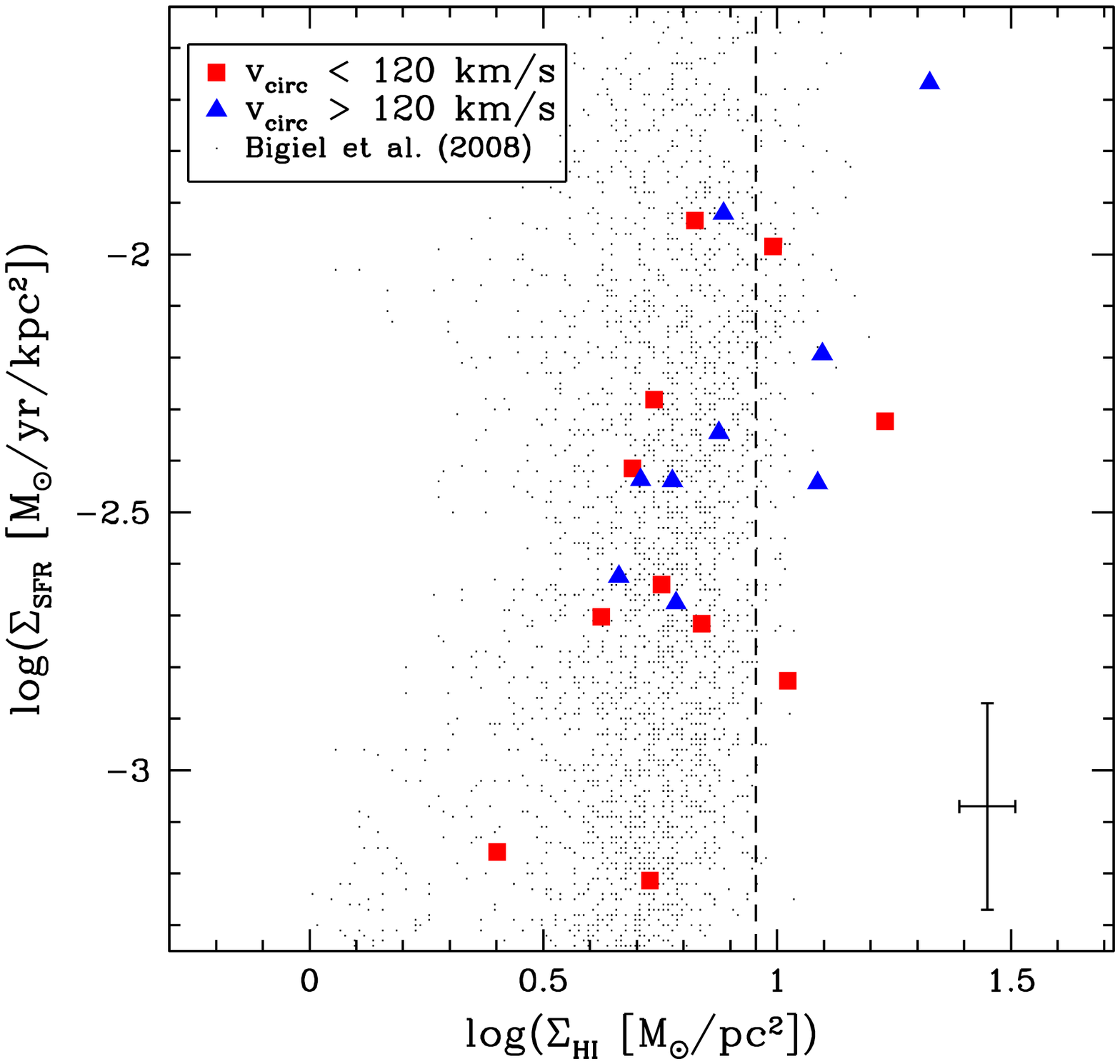}
\caption{SFR surface density versus atomic hydrogen surface density.
  Red squares represent galaxies with $v_{\rm circ} < 120 \, {\rm km
  \, s^{-1}}$.  Blue triangles represent galaxies with $v_{\rm circ} >
  120 \, {\rm km \, s^{-1}}$.  The circular velocities are from
  \ion{H}{1} rotation curve fits (Paper~I).  A representative error
  bar is shown in the lower right corner (see
  Sections~\ref{sec:gas_SD} and \ref{sec:SFR_SD} for details).  The
  vertical dashed line is at $9 \, M_{\odot} \, {\rm pc}^{-2}$, the
  typical maximum density for atomic hydrogen.  For comparison, the
  small black points represent measurements from \citet{bigiel08} from
  seven spiral galaxies sampled at $750 \, {\rm pc}$ resolution.}
\label{fig:schmidt_HI}
\end{center}
\end{figure}

We also calculated the mid-plane pressure ($P_{\rm h}$) with the
following equation from \citet{elmegreen89}:
\begin{equation}
\label{eqn:Ph}
P_{\rm h} \approx \frac{\pi}{2} \, G \, \Sigma_{\rm gas} (\Sigma_{\rm
  gas} + \frac{\sigma_{\rm gas}}{\sigma_{\ast, \rm z}} \,
  \Sigma_{\ast})
\end{equation}
We propagated the uncertainties in $l_{\ast}$, $\Sigma_{\rm gas}$, and
$\Sigma_{\ast}$ to derive a typical uncertainty in $P_{\rm h}$ of
40\%.

\section{Results}
\label{sec:results}

\subsection{Bulgeless Disk Galaxies on the Kennicutt-Schmidt Law}
\label{sec:SF_relation}
In this section, we determine whether our galaxy sample follows the
various versions of the Kennicutt-Schmidt law, i.e. the relation
between the surface density of gas (atomic, molecular, or the sum of
both) and SFR (Sections~\ref{sec:atomic_relation} -
\ref{sec:total_relation}).  In Section~\ref{sec:no_trans}, we use
these results to show that there is no transition in SFE at a circular
velocity of $120 \, {\rm km \, s^{-1}}$, or at any other circular
velocity probed by our sample ($v_{\rm circ} = 46 - 190 \, {\rm km \,
  s^{-1}}$).  The circular velocities were derived from \ion{H}{1}
rotation curve fits in Paper~I and we include them in
Table~\ref{tab:sample} for convenience.

\subsubsection{Atomic Hydrogen Kennicutt-Schmidt Law}
\label{sec:atomic_relation}
Figure~\ref{fig:schmidt_HI} shows the relationship between the star
formation rate surface density and the atomic hydrogen surface density
within the $21\arcsec$-diameter circular aperture, which corresponds
to physical diameters of 0.7 - $3.2 \, {\rm kpc}$.  Red squares denote
galaxies with $v_{\rm circ} < 120 \, {\rm km \, s^{-1}}$ and blue
triangles denote galaxies with $v_{\rm circ} > 120 \, {\rm km \,
s^{-1}}$.  The vertical dashed line is at $9 \, M_{\odot} \, {\rm
pc}^{-2}$ and represents the typical maximum \ion{H}{1} surface
density observed in most nearby galaxies \citep{bigiel08}.  For
comparison, we have shown as small dots the $750 \, {\rm pc}$-diameter
regions from the seven spiral galaxies studied in \citet{bigiel08} and
\citet{leroy08}.  Consistent with our assumptions, \citet{bigiel08}
used the same Kroupa-type IMF and do not include He in their gas
surface densities.  An important difference between our datasets is
that \citet{bigiel08} derived their SFR surface densities from a
combination of FUV (1350 - 1750$ \, {\rm \AA}$) and $24 \, \mu {\rm
m}$ data while we use H$\alpha$ and PAH emission.

The Spearman rank correlation coefficient for our data is 0.5.  For
comparison, the coefficient for the \citet{bigiel08} data is 0.4.  Our
data are more correlated than the \citet{bigiel08} data, primarily
because our sample includes several galaxies with large \ion{H}{1}
surface densities.

\begin{figure}
\epsscale{1.2}
\begin{center}
\plotone{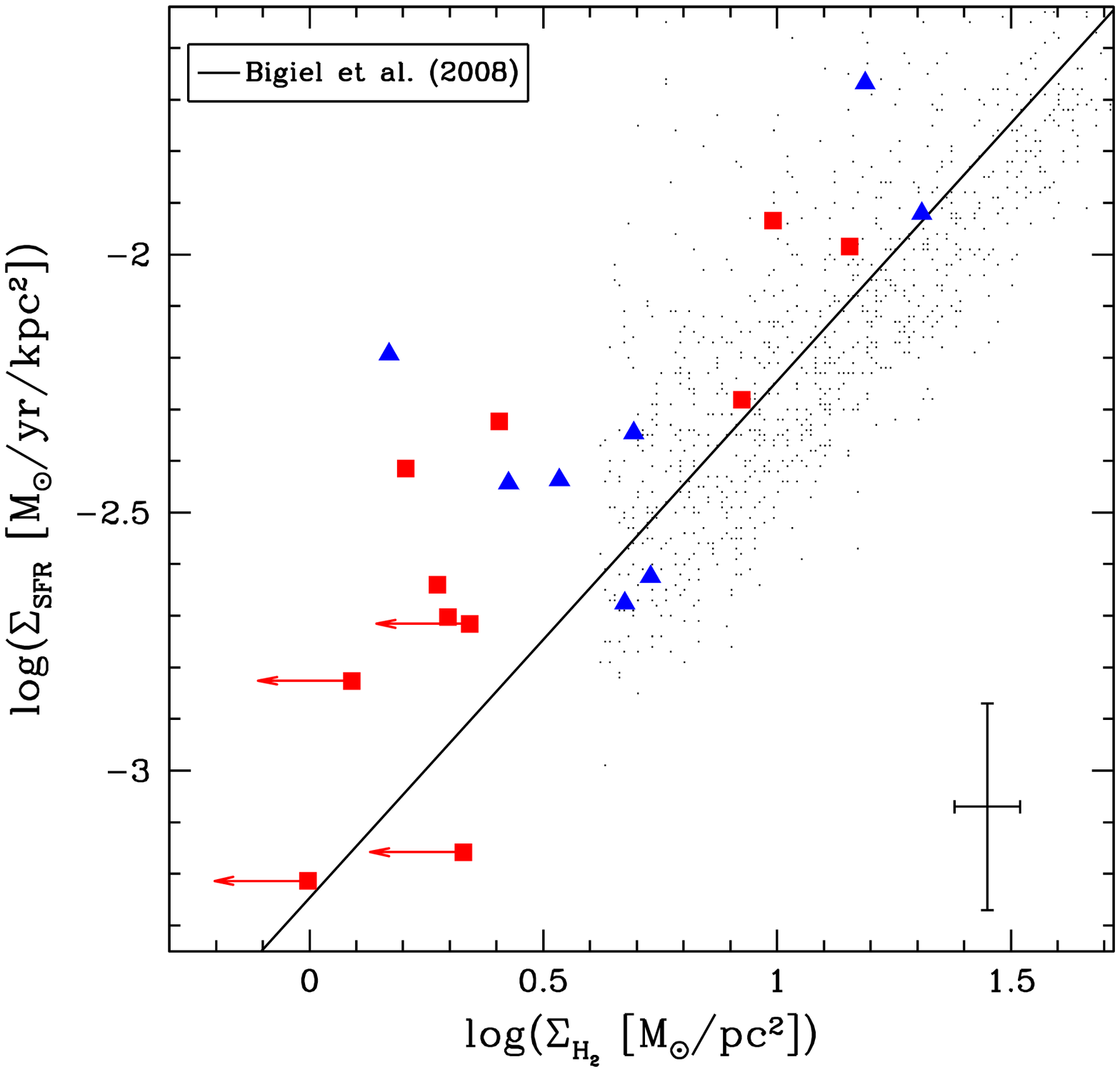}
\caption{SFR surface density versus molecular hydrogen surface
  density.  Symbols are as in Figure~\ref{fig:schmidt_HI}.  The solid
  line shows the \citet{bigiel08} fit to the molecular hydrogen star
  formation law: $\Sigma_{\rm SFR} \propto \Sigma_{\rm H_{2}}$.  We
  discuss the offset of our data relative to this fit in
  Section~\ref{sec:mol_relation}.}
\label{fig:schmidt_H2}
\end{center}
\end{figure}

\subsubsection{Molecular Hydrogen Kennicutt-Schmidt Law}
\label{sec:mol_relation}
Figure \ref{fig:schmidt_H2} shows the relationship between the star
formation rate surface density and the molecular hydrogen surface
density.  The symbols are as in Figure~\ref{fig:schmidt_HI}.  We
convert the \citet{bigiel08} $\Sigma_{\rm H_{2}}$ values to $X_{\rm
CO} = 2.8 \times 10^{20} \, {\rm cm^{-2} \, (K \, km \, s^{-1})}^{-1}$
to match our assumptions.  The solid line shows the \citet{bigiel08}
fit, also converted to the above $X_{\rm CO}$.  The Spearman rank
correlation coefficient for our data is 0.7, including upper limits.
For comparison, the coefficient for the \citet{bigiel08} data is 0.8.

Our sample of bulgeless disk galaxies, both the low- and high-$v_{\rm
circ}$ objects, appears to lie offset to higher star formation rate
surface densities relative to the \citet{bigiel08} fit.  In what
follows, we investigate the significance of and possible reasons for
this offset.

Under our assumptions (He not included in the gas surface densities,
$X_{\rm CO} = 2.8 \times 10^{20} {\rm cm^{-2} \, [K \, km \,
s^{-1}]^{-1}}$, and the Kroupa-type IMF), the \citet{bigiel08} fit is
${\rm log} \, (\Sigma_{\rm SFR}$ $[M_{\odot} \, {\rm yr^{-1} \,
kpc^{-2}}]) = a + b\, {\rm log} \, (\Sigma_{\rm H_{2}} \, [M_{\odot}
\, {\rm pc^{-2}}])$ with $a=-3.3 \pm 0.2$ and $b=1.0 \pm 0.2$.  We fit
for the intercept of our data assuming the Bigiel slope of $b=1.0$ and
including our upper limits in the fit and find $a=-3.0$ with an rms of
$0.3 \, {\rm dex}$.

We determine the significance of the offset between our sample of
bulgeless disk galaxies and the \citet{bigiel08} fit to the molecular
hydrogen star formation law by randomly selecting nineteen
measurements from \citet{bigiel08}, allowing multiple selections of
the same point \citep{press92}.  We assume a slope of 1.0, calculate
the intercept, and repeat this process $10^{6}$ times.  We find that
the probability of measuring a intercept greater than or equal to
$-3.0$ is $4 \times 10^{-6}$.

We can exclude two possible reasons for the offset.  First, the offset
is not likely due to the measurements being central values because we
confirmed that the centers of the \citet{bigiel08} galaxies are not
offset from the general trend (this can also be seen in Figure~10 of
Bigiel et al. 2008).  Second, our assumption of a single CO-to-${\rm
H}_{2}$ conversion factor probably does not lead to underestimated
molecular surface densities because our sample includes only one
galaxy for which $X_{\rm CO}$ may be underestimated because of a low
oxygen abundance, as discussed in Section~\ref{sec:gas_SD}.
Furthermore, we do not find that galaxies with lower stellar mass, and
by implication lower-$(O/H)$, are more offset from the
\citet{bigiel08} fit.  Note, however, that \citet{schruba11} did
observe that lower-oxygen abundance (down to $12+{\rm log}(O/H) =
8.25$) galaxies are offset to higher molecular SFE in their study of
33 galaxies with directly-measured oxygen abundances.  They discussed
that more observations are needed to determine if the offset is due to
$X_{\rm CO}$ variation or if it represents true SFE variation.

The quoted \citet{bigiel08} intercept uncertainty ($0.2 \, {\rm dex}$)
takes into account variations in star formation tracers, uncertainty
introduced by estimating the CO(1--0) line intensity from CO(2-1)
data, and scatter in the data.  The first of these is particularly
relevant in comparing our datasets because we trace star formation
with H$\alpha$ and PAH data and \citet{bigiel08} trace star formation
with FUV and $24 \, \mu {\rm m}$ data.  Our intercept is larger than
the \citet{bigiel08} intercept by only $1.2\sigma$, if we take their
error bar as $\sigma$.  In summary, our data are significantly offset
from the \citet{bigiel08} fit to the molecular hydrogen star formation
law in terms of statistical uncertainties, but are nearly consistent
if systematic errors are included.

\citet{kennicutt07} discussed that offsets are expected between star
formation laws derived from observations at different spatial
resolution if the power-law index of the star formation law is not
equal to one ($\Sigma_{\rm SFR} \propto \Sigma_{\rm gas}^{N}$ with $N
\ne 1$).  In transitioning from high to lower resolution, the SFR and
gas surface density will be decreased by approximately the same
factor.  If $N=1$, the lower resolution data will still lie on the
same relation as the higher-resolution data, but at lower gas and SFR
surface densities.  In contrast, if $N > 1$ ($N < 1$), the lower
resolution observations will be positively (negatively) offset from
the higher resolution observations.  This effect could contribute to
our observed offset if $N>1$ because our data probe up to $3.2 \, {\rm
kpc}$ scales and the \citet{bigiel08} data has $750 \, {\rm pc}$
resolution.  We cannot provide a firm explanation for the offset
between our data and the \citet{bigiel08} fit, but possible reasons
for the offset include the use of different star formation tracers and
resolutions, $X_{\rm CO}$ variation, and true SFE differences.

\subsubsection{Total Hydrogen Kennicutt-Schmidt Law}
\label{sec:total_relation}
\begin{figure*}
\epsscale{0.78}
\begin{center}
\plotone{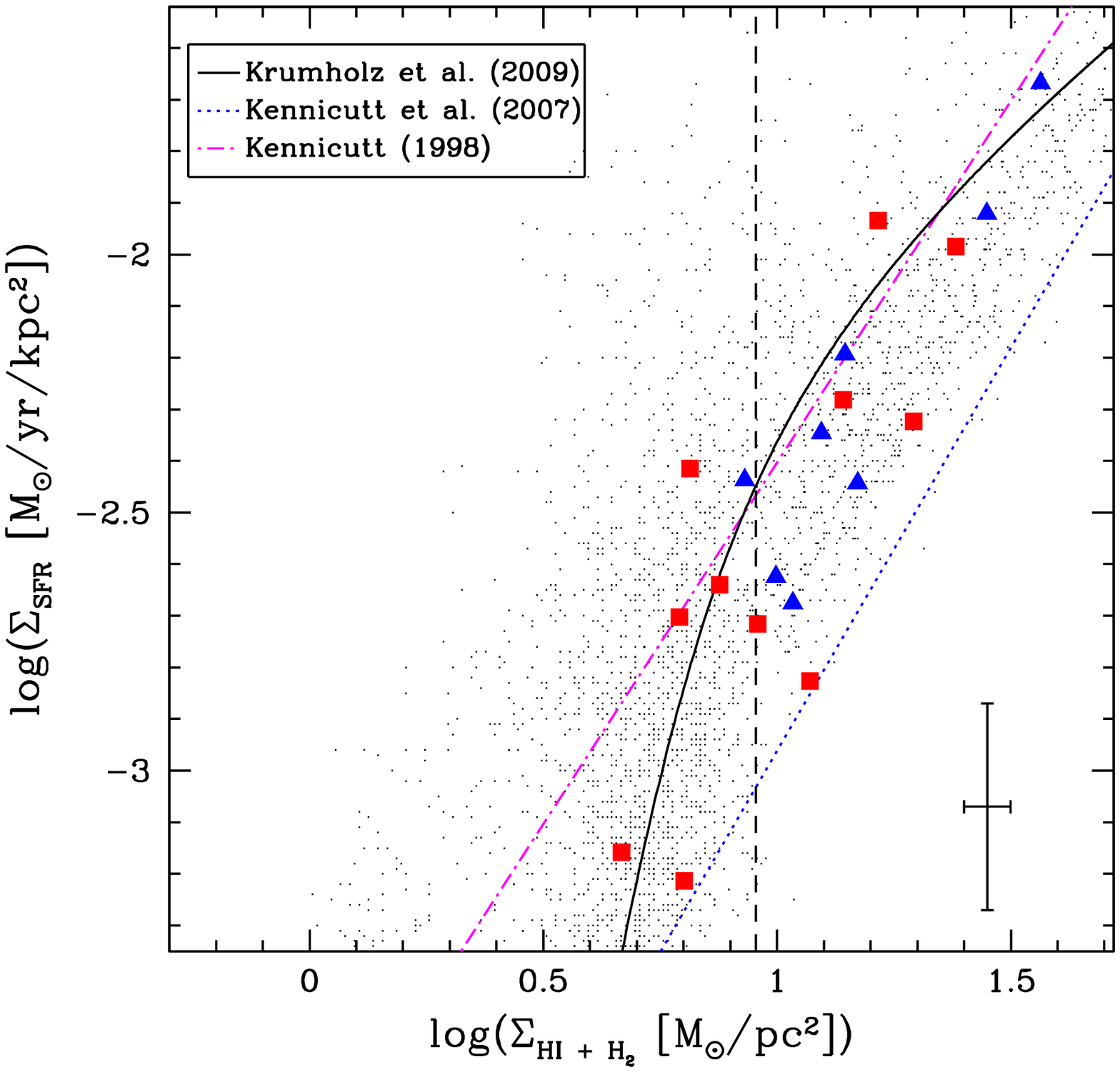}
\caption{SFR surface density versus total hydrogen surface density.
  Symbols are as in Figure~\ref{fig:schmidt_HI}.  The magenta
  dot-dashed line shows the \citet{kennicutt98} fit to a sample of 61
  normal spirals and 36 starburst galaxies, where the measurements are
  averages over the optical disk.  The dotted line shows the
  \citet{kennicutt07} fit to M51 data at $520 \, {\rm pc}$ resolution.
  The solid black line shows the \citet{krumholz09} model at $0.4 \,
  Z_{\odot}$ metallicity.}
\label{fig:schmidt_total}
\end{center}
\end{figure*}

Figure~\ref{fig:schmidt_total} shows the relationship between the star
formation rate surface density and the total hydrogen surface density.
The vertical dashed line again shows the typical maximum \ion{H}{1}
density that is observed in most nearby galaxies.  The Spearman rank
correlation coefficient for our data and also the \citet{bigiel08}
data is 0.8.  This correlation is stronger than the correlation
between SFR surface density and either atomic or molecular hydrogen
surface density.  In their study of star formation in the
atomic-dominated regime, \citet{schruba11} found that the rank
correlation coefficient between SFR surface density and total hydrogen
surface density is similar to or somewhat larger than the value for
the correlation with the molecular hydrogen surface density.  However,
they noted that data can be correlated with a strong rank statistic
even if the parameters are related by different functions in the
atomic- and molecular-dominated regimes.  As in the Schruba et
al. study, our stronger total hydrogen correlation is not likely
related to fundamental physics.

The solid black line shows the \citet{krumholz09} model with $0.4 \,
Z_{\odot}$ metallicity.  The model assumes a clumping factor, $c$,
that is the inverse of the filling factor of $\sim 100 \, {\rm
pc}$-sized atomic-molecular complexes in the beam.  This value is not
constrained well by data, so we assume the same value as
\citet{krumholz09}: $c=5$.  The magenta dot-dashed line shows the
\citet{kennicutt98} fit to a sample of normal and starburst galaxies,
where the measurements are averages over the entire optical disk.  The
blue dotted line shows the \citet{kennicutt07} fit to regions in M51
that were studied at $520 \, {\rm pc}$ resolution.
\citet{kennicutt07} attribute the offset between the blue and magenta
lines to dilution effects when the power-law index of the star
formation law is not 1.0 (as discussed in
Section~\ref{sec:mol_relation}).

\subsection{No Transition in SFE with Circular Velocity}
\label{sec:no_trans}
In Section~\ref{sec:intro}, we discussed that there could be a SFE
transition in bulgeless disk galaxies at $v_{\rm circ} = 120 \, {\rm
km \, s^{-1}}$, depending on the star formation model assumed.  In
this section we show that there is no offset between the low- and
high-$v_{\rm circ}$ galaxies on the star formation law.  In
Section~\ref{sec:discussion}, we interpret this result to discuss the
relationship between SFE and the scale height of the cold ISM.

In Figures~\ref{fig:schmidt_HI}, \ref{fig:schmidt_H2}, and
\ref{fig:schmidt_total}, there is a slight trend for low-${v_{\rm
circ}}$ galaxies to lie at lower surface densities compared to
high-$v_{\rm circ}$ galaxies, with average low- versus high-$v_{\rm
circ}$ surface densities as follows: ${\rm log} \langle \Sigma_{\rm
HI} \rangle = 0.9$ versus 1.0, ${\rm log} \langle \Sigma_{\rm H_{2}}
\rangle < 0.6$ versus $= 0.9$, ${\rm log} \langle \Sigma_{\rm HI+H_{2}}
\rangle = 1.1$ versus 1.2, and ${\rm log} \langle \Sigma_{\rm SFR}
\rangle = -2.4$ versus -2.2.  However, the differences are not
significant; the Kolmogorov-Smirnov (K-S) test probabilities that the
low- versus high-$v_{\rm circ}$ galaxies are drawn from the same
population in $\Sigma_{\rm H_{2}}$, $\Sigma_{\rm HI+H_{2}}$, and
$\Sigma_{\rm SFR}$ are 0.04, 0.2, and 0.2, respectively.

\begin{figure}
\epsscale{1.2}
\begin{center}
\plotone{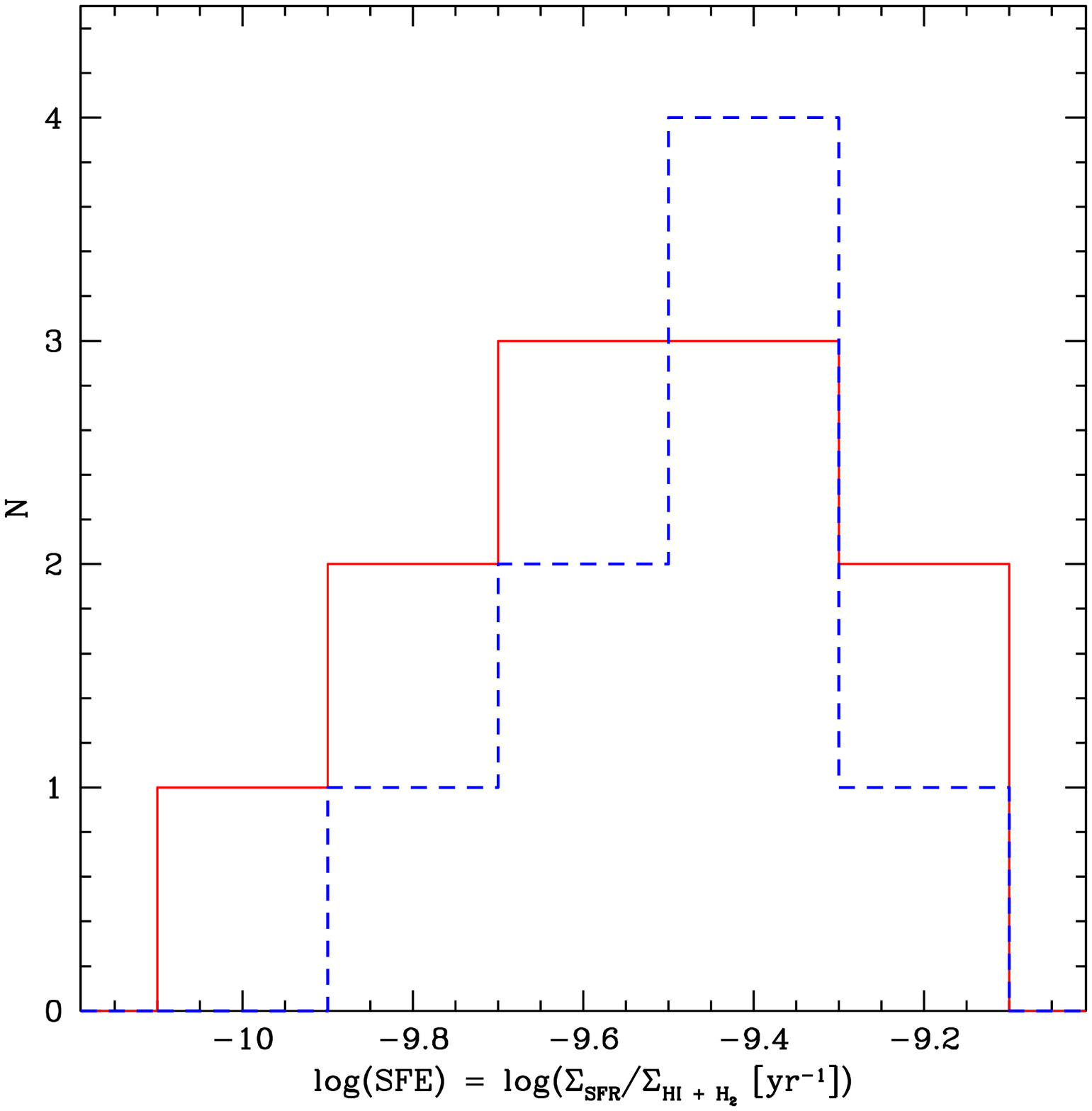}
\caption{SFE histograms of the low-$v_{\rm circ}$ (red solid) and
  high-$v_{\rm circ}$ (blue dashed) samples.  There is no significant
  difference between the distributions.}
\label{fig:SFE_hist}
\end{center}
\end{figure}

Figure~\ref{fig:SFE_hist} shows the distribution of SFE in the
low-$v_{\rm circ}$ (red solid) and high-$v_{\rm circ}$ (blue dashed)
galaxies.  If the slope of the star formation law is not 1.0, a sample
that follows the relation would have SFE that varies as a function of
$\Sigma_{\rm HI+H_{2}}$, which will tend to broaden the distributions.
However, our low- and high-$v_{\rm circ}$ objects have similar
$\Sigma_{\rm HI+H_{2}}$ distributions, so this should not lead to
spurious offsets between the samples.  Furthermore, the $\Sigma_{\rm
HI+H_{2}}$ range is not so great that it would hide any differences
between the samples.  We find no significant difference between the
SFE distributions, with a K-S test probability of 0.8 that the low-
and high-$v_{\rm circ}$ galaxies are drawn from the same population.

\begin{figure}
\epsscale{1.2}
\begin{center}
\plotone{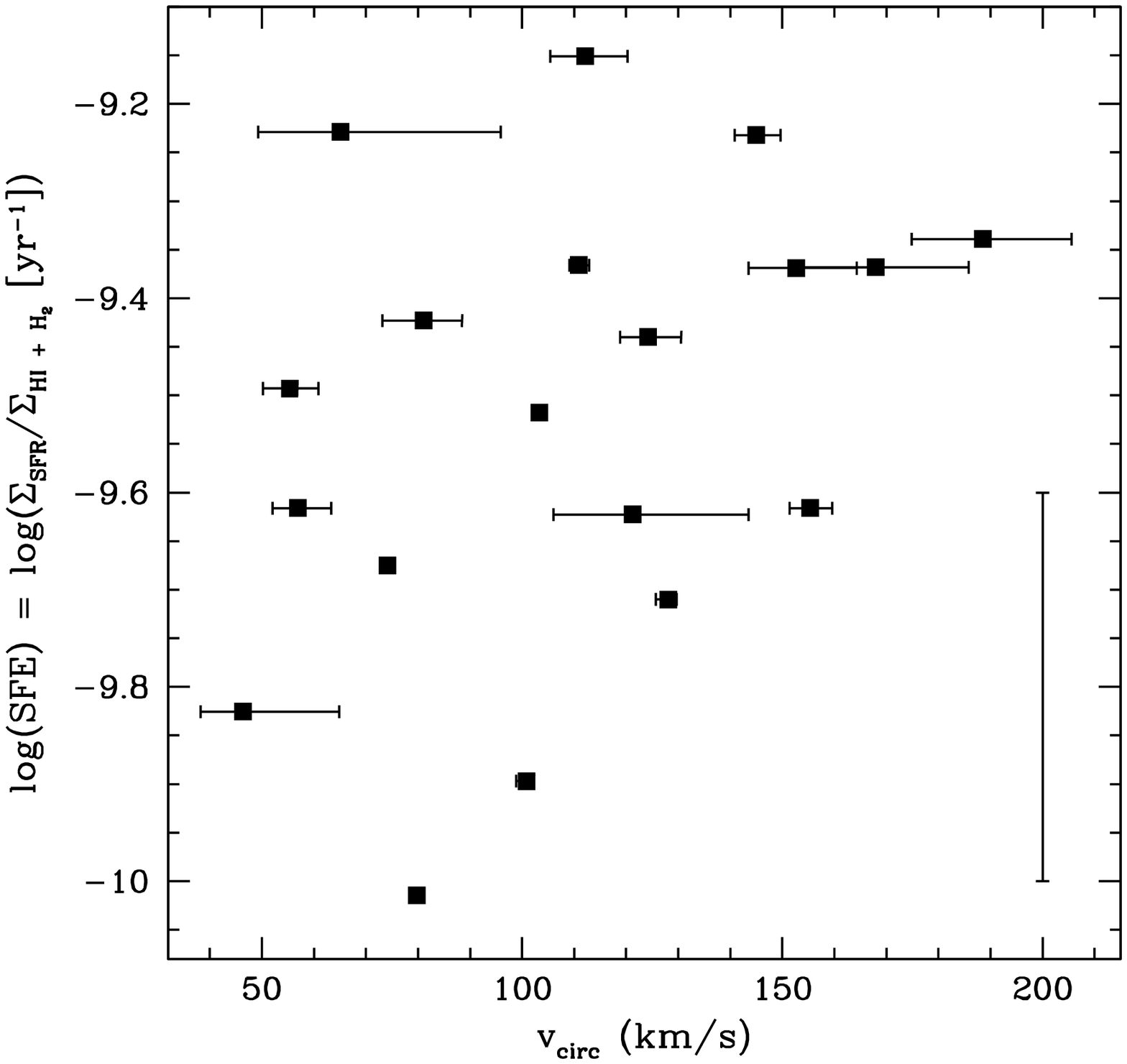}
\caption{SFE versus circular velocity.  There is no circular velocity
  at which a transition in SFE occurs.}
\label{fig:SFE_vcirc}
\end{center}
\end{figure}

\begin{figure}
\epsscale{1.2}
\begin{center}
\plotone{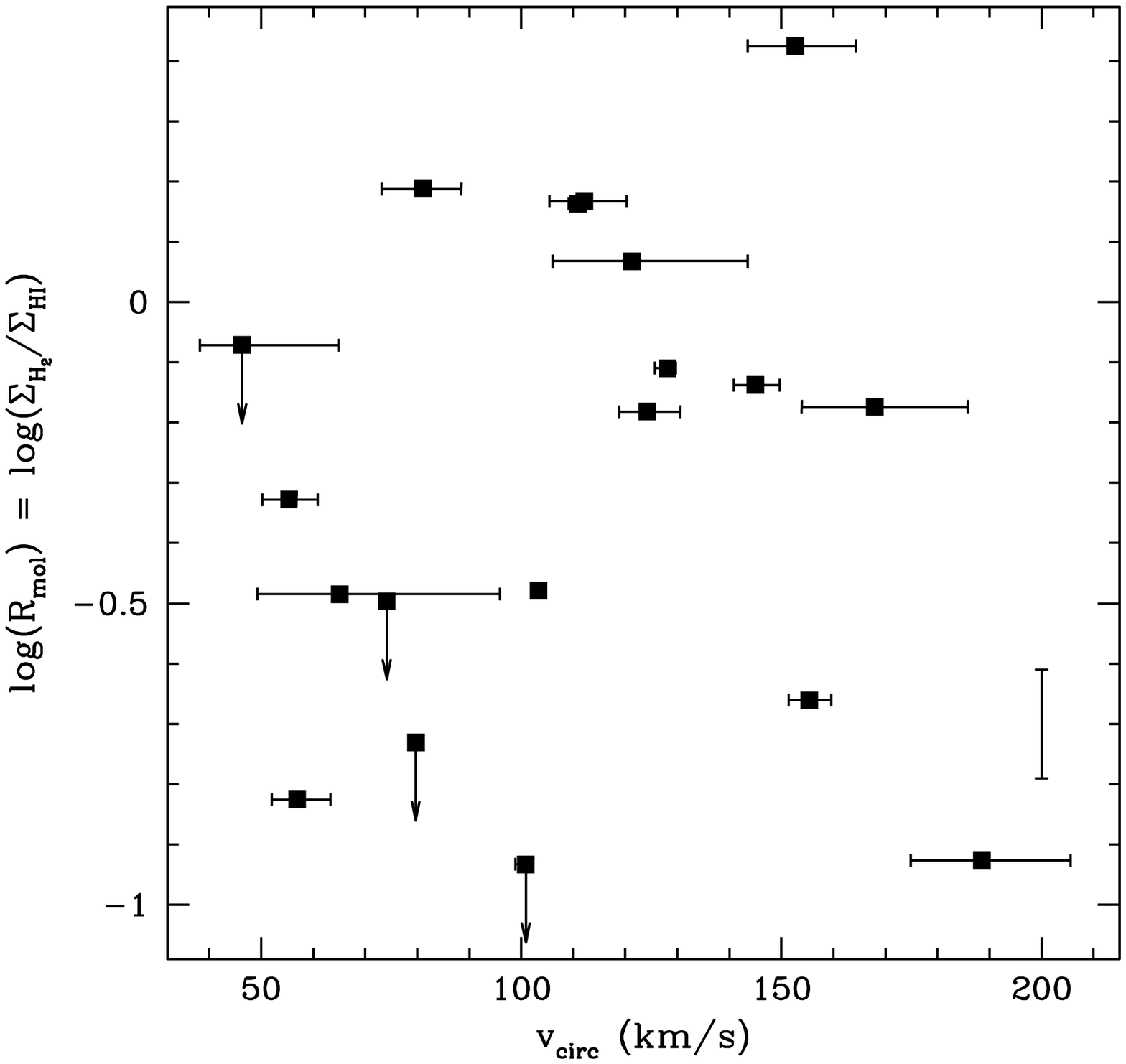}
\caption{Ratio of the molecular to atomic gas surface density ($R_{\rm
  mol}$) versus circular velocity.  There is no circular velocity at
  which a transition in $R_{\rm mol}$ occurs.}
\label{fig:Rmol_vs_vcirc}
\end{center}
\end{figure}

We conclude that our sample of bulgeless disk galaxies does not show a
strong transition in SFE at $v_{\rm circ} = 120 \, {\rm km \,
s^{-1}}$, where \citet{dalcanton04} concluded that there is a strong
transition in dust scale height.  In Figure~\ref{fig:SFE_vcirc} we
show that there is no circular velocity at which there is a transition
in SFE.  Furthermore, there is no strong trend of SFE with circular
velocity within the range we probe.  Finally, we find no difference in
the ratio of the molecular to atomic surface density between the low-
and high-$v_{\rm circ}$ galaxies (the K-S test probability that the
samples are drawn from the same $R_{\rm mol}$ population is 0.4), nor
do we find a transition in $R_{\rm mol}$ at any circular velocity
(Figure~\ref{fig:Rmol_vs_vcirc}).

\subsection{SFE Trends with Stability and Mid-plane Pressure}
\label{sec:stability}
\begin{figure*}
\epsscale{1.1}
\begin{center}
\plotone{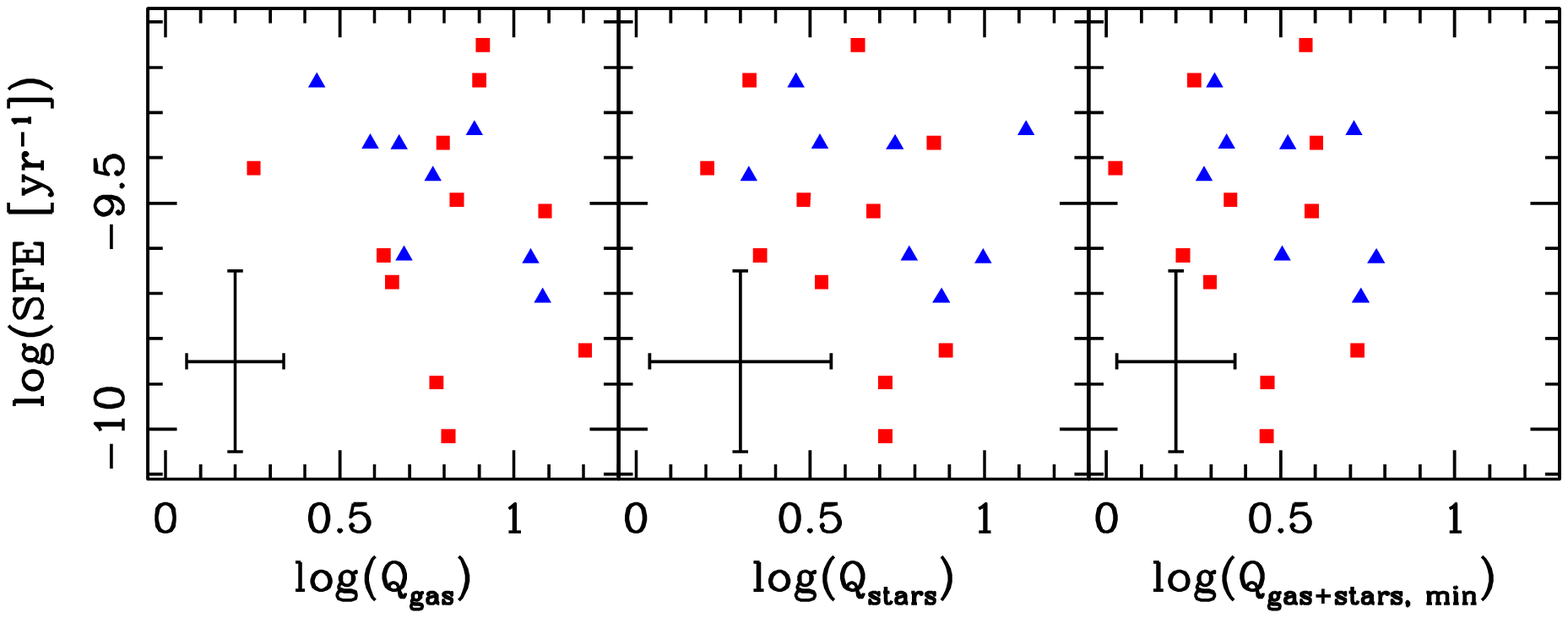}
\caption{SFE ($\Sigma_{\rm SFR} / \Sigma_{\rm HI+H_{2}}$) versus
  stability parameters $Q_{\rm gas}$ (left), $Q_{\rm stars}$ (middle),
  and $Q_{\rm gas+stars, \, min}$ (right).  The symbols are as in
  Figure~\ref{fig:schmidt_HI}.  Note that the condition for
  instability is $Q \lesssim 1$, so none of our galaxies are strictly
  unstable.  We describe our derivation of the errorbar in
  Sections~\ref{sec:SFR_SD} and \ref{sec:Q_calc}.}
\label{fig:SFE_Q}
\end{center}
\end{figure*}

In this section, we address whether our sample shows a transition or
trends in stability or mid-plane pressure, using the $Q_{\rm gas}$,
$Q_{\rm stars}$, $Q_{\rm gas+stars, min}$, and $P_{\rm h}$ values
calculated in Section~\ref{sec:Q_calc}.  Figure~\ref{fig:SFE_Q} shows
the SFE versus $Q_{\rm gas}$ (left), $Q_{\rm stars}$ (middle), and
$Q_{\rm gas+stars, min}$ (right).  We find no unstable regions.  The
SFE generally decreases with larger, more stable Q values, although
the correlation is not strong.  For SFE versus $Q_{\rm gas}$, $Q_{\rm
stars}$, and $Q_{\rm gas+stars, min}$, we find Spearman rank
correlation coefficients of -0.2, -0.3, and -0.2, respectively.  The
sign of this correlation is as expected, but one might instead have
expected a sharp decrease in SFE when Q rises above 1.  These results
are qualitatively similar to those in \citet{leroy08}.

\citet{dalcanton04} concluded that galaxies with ${\rm v_{circ}} < 120
\, {\rm km \, s^{-1}}$ are generally stable while galaxies with ${\rm
v_{circ}} > 120 \, {\rm km \, s^{-1}}$ are generally unstable,
especially in the central ($r<l_{\ast}$) regions.  Contrary to these
results, we see no evidence for a transition in stability at ${\rm
v_{circ}} = 120 \, {\rm km \, s^{-1}}$; the K-S test probability that
the low- and high-$v_{\rm circ}$ galaxies are drawn from the same
population of $Q_{\rm gas}$, $Q_{\rm stars}$, and $Q_{\rm gas+stars,
min}$ is 0.5, 0.2, and 0.8, respectively.

There are two principal differences between our assumptions for the
stability inputs and those in \citet{dalcanton04}.  First, we use a
constant gas velocity dispersion and they used the quadrature sum of
the velocity dispersion of atomic and molecular gas (10 and $5 \, {\rm
km \, s^{-1}}$, respectively), weighted by the relative mass surface
densities of the components.  We use these assumptions and still find
no difference in stability between the low- and high-$v_{\rm circ}$
galaxies.  Second, \citet{dalcanton04} estimated the molecular
hydrogen surface densities for their sample from a scaling with
circular velocity ($\Sigma_{\rm H_{2}} = (v_{\rm circ}/47.1 \, {\rm km
\, s^{-1}})^{2.49}$).  This scaling relation was derived from a
similar sample of galaxies with $\Sigma_{\rm H_{2}}$ values from
\citet{rownd99} and $v_{\rm circ}$ estimated from single-dish
\ion{H}{1} data.  Molecular hydrogen surface densities calculated
using the scaling with $v_{\rm circ}$ are too large by a factor of six
for our sample of high-$v_{\rm circ}$ galaxies and too large by a
factor of two for our sample of low-$v_{\rm circ}$ galaxies.  This
discrepancy leads to smaller, less stable $Q_{\rm gas}$ and $Q_{\rm
gas+stars}$ in the high-$v_{\rm circ}$ galaxies relative to the
low-$v_{\rm circ}$ galaxies.  However, even under these assumptions
there is no significant difference between the low- and high-$v_{\rm
circ}$ galaxies in their $Q$ values. Note that the difference in our
molecular hydrogen surface densities compared to the values derived
from the scaling with $v_{\rm circ}$ may also be related to resolution
differences.  Our $\Sigma_{\rm H_{2}}$ measurements are within the
central $21\arcsec$ while the \citet{rownd99} data have $45\arcsec$
resolution and a similar distance distribution.

\begin{figure}
\epsscale{1.2}
\begin{center}
\plotone{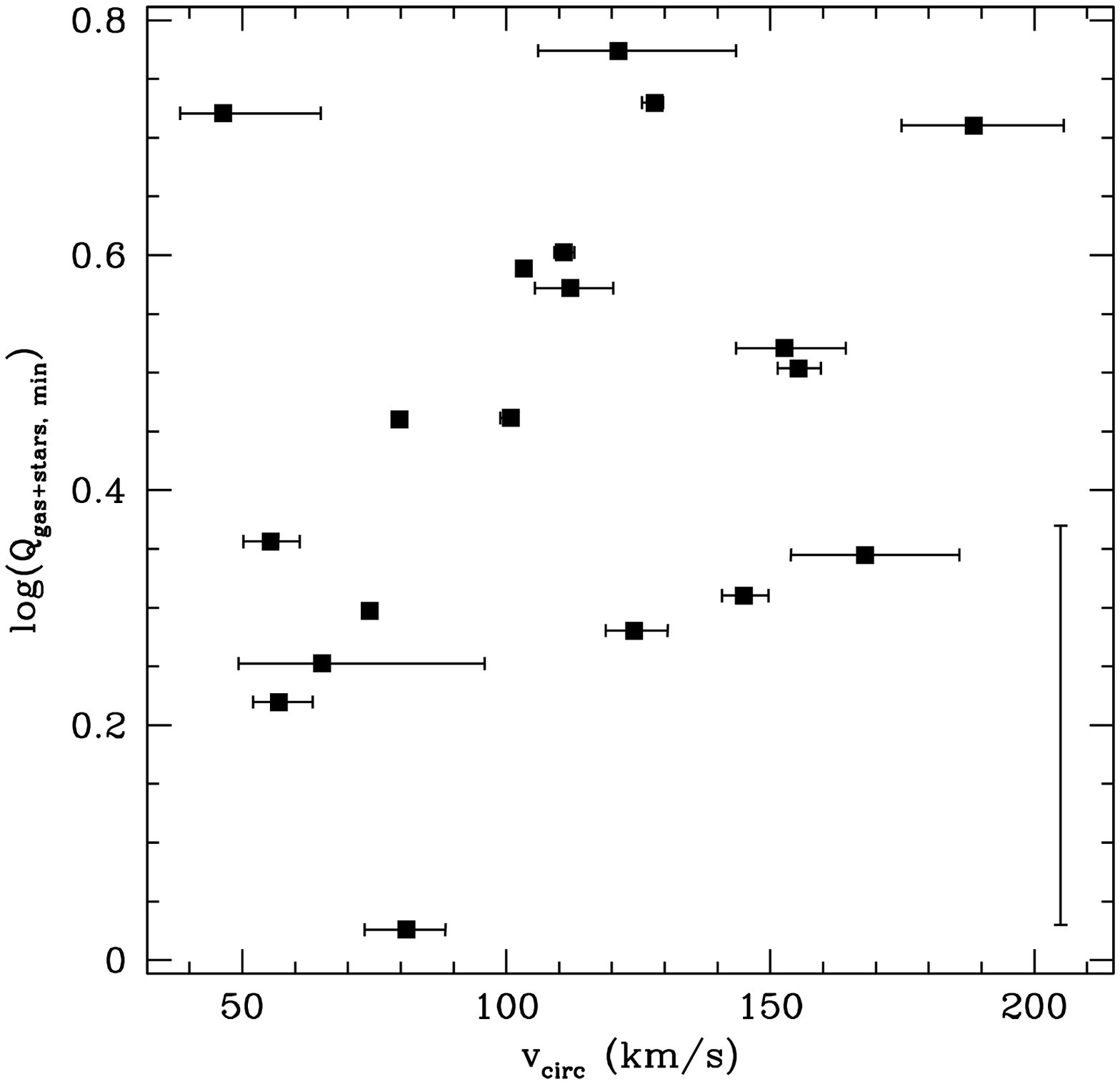}
\caption{Stability parameter, including the gas and stellar
    contribution, versus circular velocity.  There is no circular
    velocity at which a transition in stability occurs.  We describe
    our derivation of the vertical errorbar in
    Section~\ref{sec:Q_calc}.}
\label{fig:Q_vcirc}
\end{center}
\end{figure}

We test whether a stability transition occurs at any circular velocity
in Figure~\ref{fig:Q_vcirc}, where we plot $Q_{\rm gas+stars, min}$
versus $v_{\rm circ}$.  There is no circular velocity below which the
regions are stable and above which the regions are unstable.  If any
trend is present, it is that higher circular velocity objects are more
stable.

\begin{figure}
\epsscale{1.2}
\begin{center}
\plotone{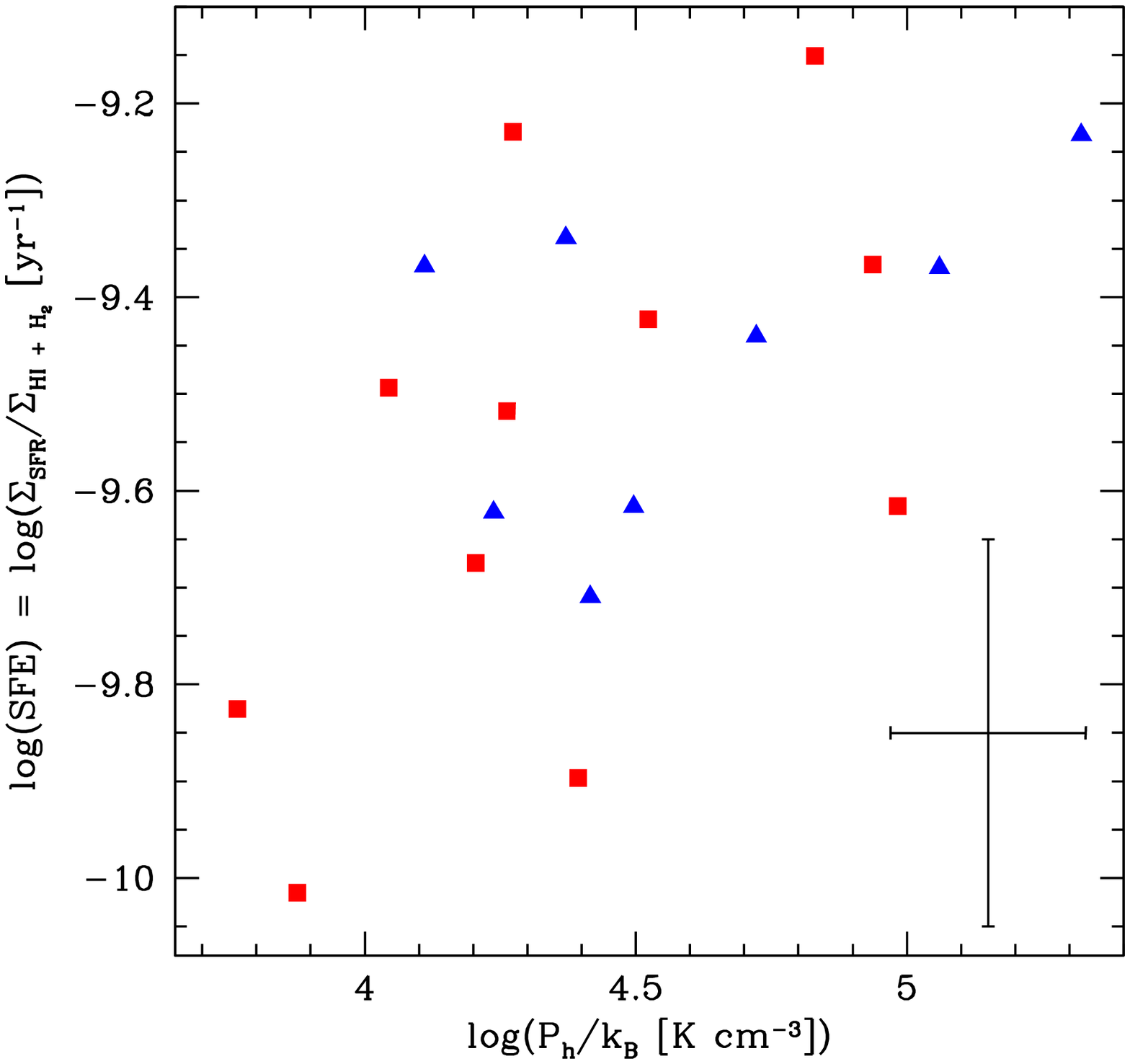}
\caption{SFE versus mid-plane pressure.  Symbols are as in
Figure~\ref{fig:schmidt_HI}.}
\label{fig:SFE_Ph}
\end{center}
\end{figure}

We plot the SFE versus the mid-plane pressure in
Figure~\ref{fig:SFE_Ph}.  As seen in \citet{leroy08}, we find an
increase in SFE with increasing mid-plane pressure, but we do not
probe high enough $P_{\rm h}$ values or have enough data points to
sample the constant-SFE region of the diagram that is clear in
\citet{leroy08}.  We find no difference between the low- and
high-$v_{\rm circ}$ galaxies in their $P_{\rm h}$ distributions, with
a K-S test probability of 0.7 that they are drawn from the same
population.  Even focusing only on galaxies with the same $\Sigma_{\rm
HI+H_{2}}$ (within the uncertainty in the parameter), the median
pressure is the same in low- and high-$v_{\rm circ}$ galaxies.

\newpage

\subsection{Dependence of Star Formation on Metallicity}
\label{sec:metallicity}
In this section, we use the oxygen abundance estimated from the
stellar mass to study how star formation depends on metallicity and
compare these results to recent theoretical work by
\citet{krumholz09}.  At a given total gas surface density, the
\citet{krumholz09} model predicts lower SFE at lower metallicity
because ${\rm H_{2}}$ survival requires a higher column density as
${\rm H_{2}}$ self-shielding becomes more important than shielding by
dust.  This metallicity dependence distinguishes the model from other
leading models, such as the model where mid-plane pressure determines
where the ISM is molecular.

\begin{figure*}
\begin{center}
\plottwo{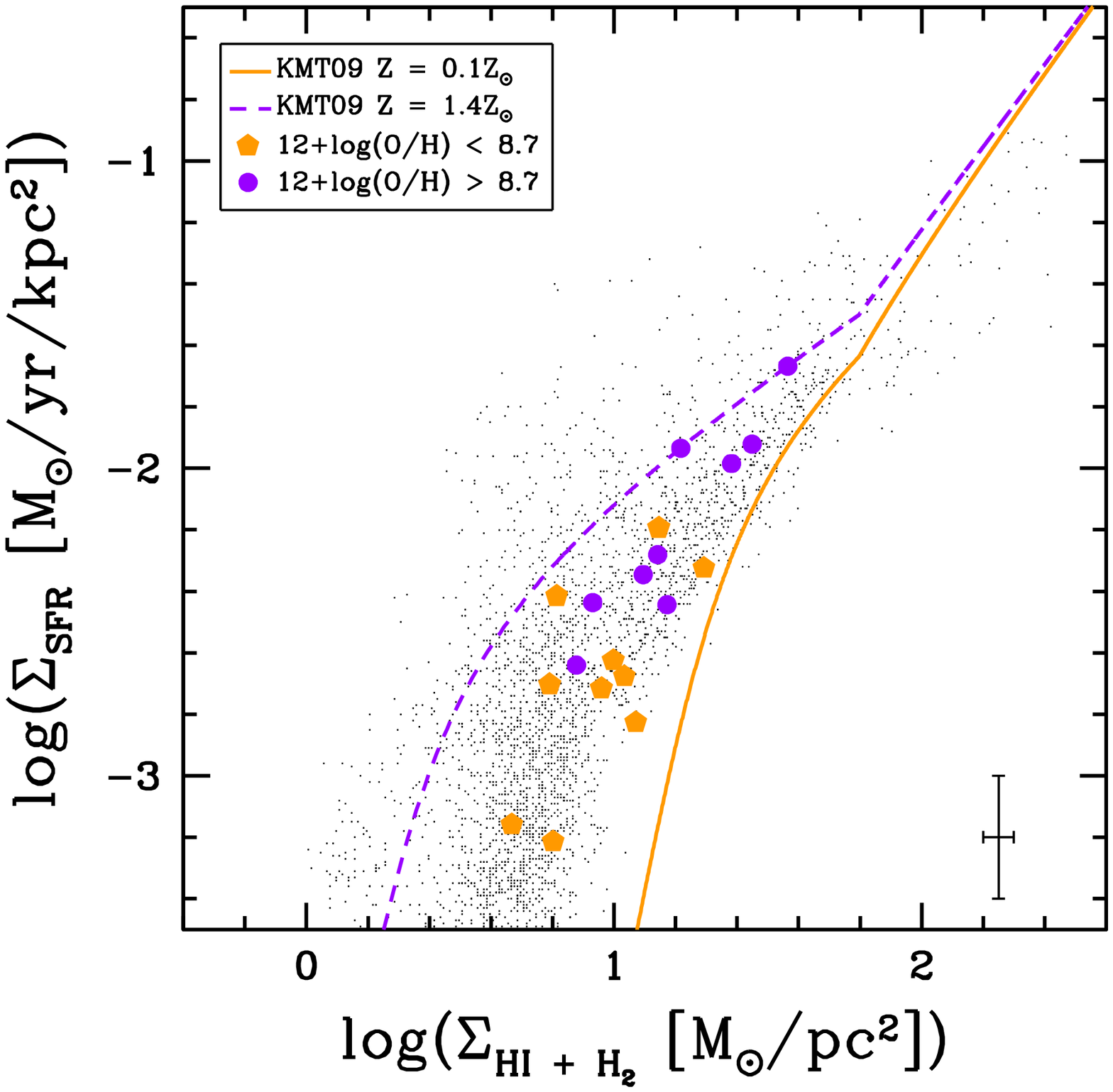}{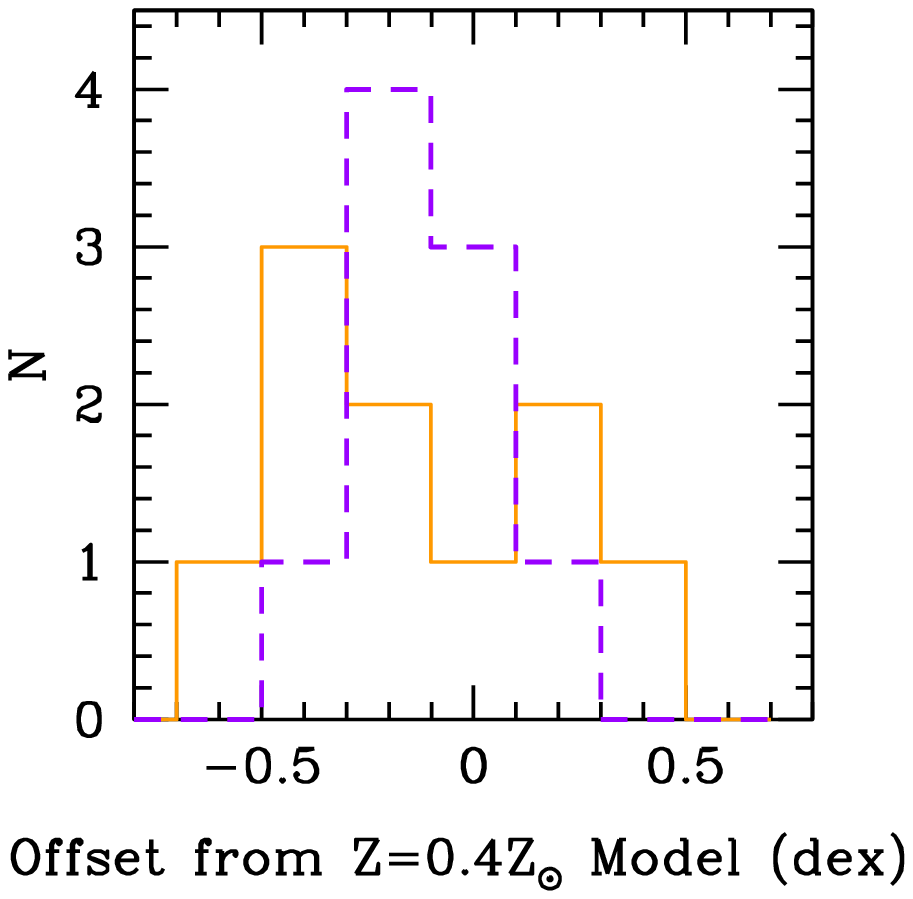}
\caption{Left: SFR surface density versus total hydrogen surface
  density, as in Figure~\ref{fig:schmidt_total}, but showing a larger
  range to demonstrate the model trends.  The data are divided by
  oxygen abundance, where orange pentagons represent galaxies with
  $12+{\rm log} \, (O/H) < 8.7$ and purple circles represent galaxies
  with $12+{\rm log} \, (O/H) > 8.7$.  The lines show the
  \citet{krumholz09} model with a metallicity of $0.1 \, Z_{\odot}$
  (orange solid) and $1.4 \, Z_{\odot}$ (purple dashed), which
  correspond to the lowest and highest metallicities in our sample.
  Right: Histograms of the offset of the data from the
  \citet{krumholz09} model with a metallicity of $0.4 \, Z_{\odot}$.
  The orange line represents the distribution of galaxies with
  $12+{\rm log} \, (O/H) < 8.7$ and the purple line represents the
  distribution of galaxies with $12+{\rm log} \, (O/H) > 8.7$.  There
  is no significant difference between the distributions.}
\label{fig:schmidt_met}
\end{center}
\end{figure*}

The left panel of Figure~\ref{fig:schmidt_met} shows the total
hydrogen star formation law, as plotted in
Figure~\ref{fig:schmidt_total}, except the galaxies are divided into
low-$(O/H)$ and high-$(O/H)$, with the division at $12+{\rm log} (O/H)
= 8.7$.  The orange solid and purple dashed lines show the model of
\citet{krumholz09} at $0.1 \, Z_{\odot}$ and $1.4 \, Z_{\odot}$,
respectively. These metallicities correspond to the extrema of our
dataset, assuming $Z/Z_{\odot} = (O/H)/(O/H)_{\odot}$ and $12+{\rm
log}(O/H)_{\odot} = 8.86$ \citep{delahaye06}.  We see no evidence that
the low- and high-$(O/H)$ galaxies cluster towards the low- and
high-$Z$ models, respectively.

The right panel of Figure~\ref{fig:schmidt_met} shows the offset of
our data from the $0.4 \, Z_{\odot}$ \citet{krumholz09} model, which
provides the best fit to our dataset as a whole (although note that we
would expect the best fit to be the $0.7 \, Z_{\odot}$ model, as that
corresponds to the average $(O/H)$ of our full sample).  The orange
solid and purple dashed lines show the distributions of our
low-$(O/H)$ and high-$(O/H)$ galaxies, as divided in the left panel.
If our data clearly followed the \citet{krumholz09} model,
high-$(O/H)$ galaxies would be positively offset from the $0.4 \,
Z_{\odot}$ model and low-$(O/H)$ galaxies would be negatively offset.
We find no significant difference between the offset of the low-
versus high-$(O/H)$ galaxies (the K-S test probability that the
samples are drawn from the same offset population is 0.14).  A second
order effect is that the offset from the $0.4 \, Z_{\odot}$ model is
expected to decrease with increasing $\Sigma_{\rm HI+H_{2}}$.  This
may be related to the narrower offset distribution for the
high-$(O/H)$ galaxies, which tend to have higher $\Sigma_{\rm
HI+H_{2}}$.

\begin{figure}
\begin{center}
\epsscale{1.2}
\plotone{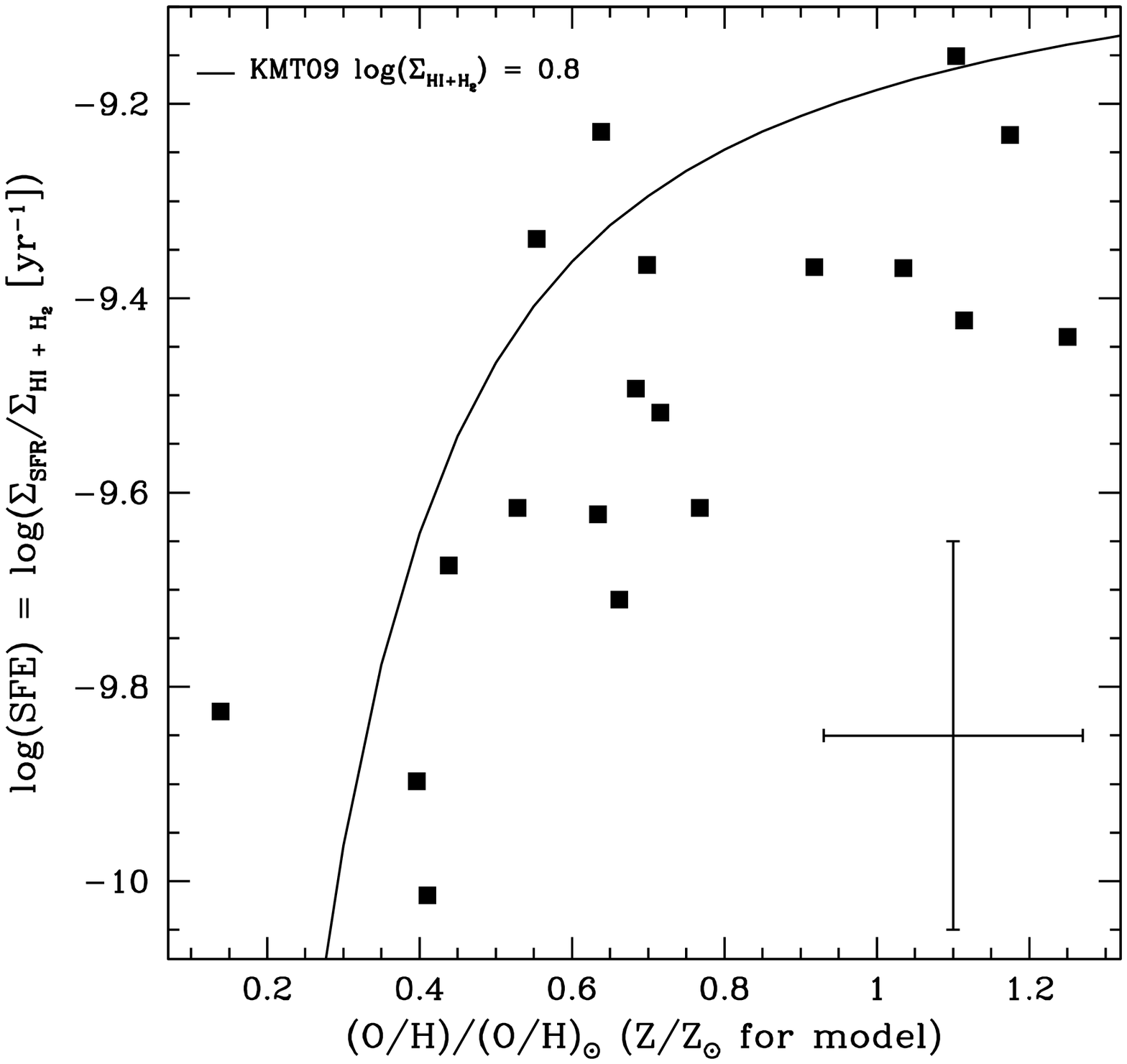}
\caption{SFE versus oxygen abundance, scaled to the solar value.  The
  solid line demonstrates the SFE trend with metallicity for the
  \citet{krumholz09} model at ${\rm log}(\Sigma_{\rm HI+H_{2}}) =
  0.8$.}
\label{fig:SFE_OH}
\end{center}
\end{figure}

Figure~\ref{fig:SFE_OH} shows that there is a correlation between SFE
and oxygen abundance in our data.  For comparison, the line represents
the \citet{krumholz09} model, where the SFE depends on metallicity,
$\Sigma_{\rm gas}$, and the filling factor.  We have plotted the
\citet{krumholz09} model with ${\rm log} \, (\Sigma_{\rm HI+H_{2}}) =
0.8$ (no He).  This is a low $\Sigma_{\rm HI+H_{2}}$ value compared to
our data.  Higher $\Sigma_{\rm HI+H_{2}}$ models reach constant SFE at
lower metallicity.  The mismatch between the average $\Sigma_{\rm
HI+H_{2}}$ in the data and the best-fit $\Sigma_{\rm HI+H_{2}}$ used
in the model is likely due to our assumptions for $X_{\rm CO}$ and the
IMF in the data, our filling factor assumption in the model, and to
uncertainties in the metallicity scale.  For the latter, various
strong-line metallicity methods return $12+{\rm log}(O/H)$ values as
disparate as $0.5 \, {\rm dex}$ \citep{kewley08}, and it is unknown
which calibration best aligns with the solar value.  Therefore, the
model line in Figure~\ref{fig:SFE_OH} only represents the expected
trend of lower SFE at lower metallicity, at a given $\Sigma_{\rm
gas}$.  At this level, the data do show the expected trend, with a
Spearman correlation coefficient of 0.6.  However, upon closer
inspection, this agreement is mainly due to the fact that
lower-$(O/H)$ galaxies tend to have lower $\Sigma_{\rm HI+H_{2}}$
(this is most obvious in Figure~\ref{fig:schmidt_met}, although also
note that the K-S test probability is 0.05 that the $\Sigma_{\rm
HI+H_{2}}$ distributions of the low- and high-$(O/H)$ samples are
drawn from the same population, which is not strong evidence for a
difference).  Because of their lower $\Sigma_{\rm HI+H_{2}}$ values,
the low-$(O/H)$ galaxies are closer to the atomic-dominated regime of
the star formation law where SFEs are lower.  This is of course
consistent with the \citet{krumholz09} model, but is not a
discriminating test of the model because many other properties
correlate with gas surface density \citep[for examples,
see][]{leroy08}.  A sample with a range of directly-measured
metallicities within a small range of gas surface density would
provide a more discriminating test.  In summary, we see no clear
evidence to support the \citet{krumholz09} model, but our data are
also not inconsistent with it.

\section{Discussion}
\label{sec:discussion}
In Section~\ref{sec:no_trans}, we found that there is no transition in
molecular fraction or SFE at any circular velocity probed by our
sample.  In Section~\ref{sec:stability}, we found that all our
galaxies are formally stable.  While we did find a general trend of
decreasing SFE with larger, more stable Q values, we found no sharp
transition in stability at any circular velocity.  Finally in
Section~\ref{sec:metallicity}, we found that SFE decreases at lower
oxygen abundance, but the trend is not a particularly constraining
test of the \citet{krumholz09} model.  In this section, we first
address our assumption that there is a transition in dust scale height
at $v_{\rm circ} = 120 \, {\rm km \, s^{-1}}$, rather than a
transition in dust content.  We then interpret our results to discuss
the relationship between SFE and the scale height of the cold ISM.
Finally, we comment on the scale of physical processes that affect
star formation and discuss our results in the context of leading star
formation models.

\subsection{A Transition in Dust Scale Height versus Dust Content}
\label{sec:discussion1}
In this section, we investigate the argument that the dust structure
transition observed by \citet{dalcanton04} is due to a transition in
dust scale height rather than due to a transition in the amount of
dust present.  \citet{dalcanton04} came to this conclusion because the
dust structure transition occurs over a narrow range in circular
velocity, where a large change in the DGR, and therefore dust content,
is unexpected.  All but one galaxy in our sample are detected at {\it
Infrared Astronomical Satellite} 60 and 100$\, \mu {\rm m}$, which
indicates that there is at least some dust in the low-$v_{\rm circ}$
galaxies.  We estimated a DGR proxy as the ratio of the total infrared
luminosity ($L_{\rm TIR}$), calculated with IRAS 25, 60, and 100$\,
\mu {\rm m}$ data \citep{moshir90} and the \citet{dale02} relation, to
the combined atomic and molecular hydrogen mass.  We find no
transition in this DGR proxy at $v_{\rm circ} = 120 \, {\rm km \,
s^{-1}}$, nor do we find any correlation between the DGR proxy and
circular velocity.  $L_{\rm TIR}$ is not necessarily proportional to
the dust mass because the dust temperature may not be the same for all
the galaxies; nevertheless, there is clearly a significant amount of
dust in the low-$v_{\rm circ}$ galaxies.

Two recent studies have found convincing evidence that low-$v_{\rm
circ}$ spirals have large dust scale heights.  \citet{seth05} carried
out a resolved stellar population study of six edge-on, late-type
spirals with $v_{\rm circ} = 67 - 131 \, {\rm km \, s^{-1}}$ and found
that the scale height of the young ($\lesssim 10^{8} \, {\rm yr}$)
stellar population, which presumably formed from the cold ISM, is
larger than in the Milky Way.  \citet{maclachlan11} modeled the
spectral energy distribution (SED) of three edge-on, low surface
brightness galaxies with $v_{\rm circ} = 88 - 105 \, {\rm km \,
s^{-1}}$ and found that a significant amount of dust must be present
to account for the FIR (70 and 160$ \, \mu {\rm m}$) emission, but the
dust must have a large scale height such that it does not
significantly obscure the optical emission.  The authors concluded
that the galaxies have dust scale heights greater than or equal to the
stellar scale heights.  This is in contrast to modeling studies of
high surface brightness galaxies, which concluded that the dust scale
height is about half the stellar scale height
\citep[e.g.,][]{xilouris99}.

\citet{hunter06} provided an alternative explanation for the dust
structure transition observed by \citet{dalcanton04}.  The authors
studied a large sample of irregular galaxies and found that the
average B-band surface brightness is smaller than in higher-mass
spiral galaxies.  Based on this result, they suggested that lower
stellar surface density is the cause of the larger scale heights in
low-$v_{\rm circ}$ galaxies.  They discussed that disk stability could
correlate with the dust structure but not be the cause of the
transition because the gas scale height \citep[$h_{\rm gas} \propto
\frac{\sigma_{\rm gas}^2}{G \, \rho}$, where $\rho$ is the mass
volume density of gas and stars;][]{blitz04} shares many of the same
parameters with the stability parameter.

The authors discussed that dust opacity may also contribute to the
observed difference in dust structure because of two effects.  First,
lower-$v_{\rm circ}$ objects should have lower metallicity, and
therefore lower dust content.  Second, the scale length of a
lower-mass galaxy is smaller and therefore a low-$v_{\rm circ}$,
edge-on galaxy will have less depth from the edge to the center over
which to accumulate dust column density compared to a high-$v_{\rm
circ}$ galaxy.  Note that no scale height transition is needed in this
interpretation.  However, dust opacity is not likely the sole cause of
the dust structure transition because of the reasons listed above.
\citet{hunter06} suggested that the observed dust structure transition
is likely due to a combination of stellar surface density and dust
opacity.  The authors agreed that a scale height transition does
occur, so if this interpretation is correct, we can still constrain
the effect of scale height differences on SFE.

In Sections~\ref{sec:SFE_scale_height} and \ref{sec:models}, we assume
that the dust scale heights in low-$v_{\rm circ}$ galaxies are a
factor of two larger than in high-$v_{\rm circ}$ galaxies.
\citet{dalcanton04} estimated this factor by examining the dust
morphology in {\it Hubble Space Telescope} images of a couple edge-on,
late-type galaxies.  This factor is consistent with the SED modeling
results of \citet{maclachlan11} and \citet{xilouris99} if the stellar
scale height distributions of low- and high-$v_{\rm circ}$ galaxies
have significant overlap, which \citet{dalcanton04} found to be true
(although there are low-$v_{\rm circ}$ galaxies where the stellar
scale height is about half the value in high-$v_{\rm circ}$ galaxies,
in which case the dust scale heights may be comparable).

\subsection{The Relationship between SFE and Scale Height}
\label{sec:SFE_scale_height}
In Section~\ref{sec:no_trans}, we found no transition in SFE at
$v_{\rm circ} = 120 \, {\rm km \, s^{-1}}$ or at any circular velocity
probed by our sample.  In this section, we interpret this result to
discuss the relationship between SFE and the scale height of the cold
ISM.  For this discussion, we make a number of assumptions.  First, we
assume that our sample of bulgeless disk galaxies is similar to that
of \citet{dalcanton04} in that galaxies with $v_{\rm circ} > 120 \,
{\rm km \, s^{-1}}$ have narrow dust lanes while galaxies with $v_{\rm
circ} < 120 \, {\rm km \, s^{-1}}$ have no obvious dust lanes.  This
is a reasonable assumption because it was our main consideration in
choosing the sample, but because our galaxies are moderately inclined
rather than edge-on, we are unable to directly measure the vertical
dust structure.  Second, as in \citet{dalcanton04}, we assume that the
dust scale heights in the low-$v_{\rm circ}$ galaxies are larger than
in the high-$v_{\rm circ}$ galaxies (Section~\ref{sec:discussion1}
addresses this assumption).  Finally, we assume that the molecular gas
and dust scale heights are comparable.

\citet{dalcanton04} suggested that there may be a SFE transition
associated with the dust scale height transition at $v_{\rm circ} =
120 \, {\rm km \, s^{-1}}$.  The authors gave an example that assumes
the true star formation law is a correlation between the volume
density of gas ($\rho_{\rm gas}$) and the SFR volume density
($\rho_{\rm SFR}$): $\rho_{\rm SFR} \propto \rho_{\rm gas}^{N}$.  The
larger scale heights of low-$v_{\rm circ}$ galaxies lead to lower gas
volume densities.  Depending on the index, $N$, a low-$v_{\rm circ}$
galaxy with the same gas surface density as a high-$v_{\rm circ}$
galaxy can have a lower SFR surface density and therefore a lower SFE.
To consider this point, we first assume the same $\Sigma_{\rm gas}$
for a low- and high-$v_{\rm circ}$ galaxy and the following
relationships between the surface and volume density of gas and SFR:
$\Sigma_{\rm gas} \propto \rho_{\rm gas} \, h$, and $\Sigma_{\rm SFR}
\propto \rho_{\rm SFR} \, h$, where $h$ is the scale height of the
star forming gas and newly formed stars and the proportionality
constant is the same for both relationships.  We set $\beta$ equal to
the ratio of dust scale heights in low-$v_{\rm circ}$ (lv) versus
high-$v_{\rm circ}$ (hv) galaxies: $\beta = h_{\rm lv}/h_{\rm
hv}$. \citet{dalcanton04} very approximately estimated this ratio to
be about two.  Under these assumptions, $\Sigma_{\rm SFR, hv} =
\beta^{N-1} \, \Sigma_{\rm SFR, lv}$, where $\Sigma_{\rm SFR, hv}$ and
$\Sigma_{\rm SFR, lv}$ are the star formation rate surface densities
of the low- and high-$v_{\rm circ}$ galaxy with the same $\Sigma_{\rm
gas}$.  \citet{dalcanton04} discussed the case where $\beta=2$ and
$N=1.5$, where we expect the high-$v_{\rm circ}$ galaxy to have a
$\Sigma_{\rm SFR}$ that is a factor of 1.4 larger than the low-$v_{\rm
circ}$ galaxy.  Note that if $N=1$ there is no expected difference in
star formation rate surface density.

In the star formation law plots of Figures~\ref{fig:schmidt_HI},
\ref{fig:schmidt_H2}, and \ref{fig:schmidt_total}, we are sensitive to
offsets in intercept between the low- and high-$v_{\rm circ}$ samples
that are greater than the uncertainty in the intercept, which is $\sim
0.3 - 0.4 \, {\rm dex}$, assuming an uncertainty in $\Sigma_{\rm SFR}$
of $0.2 \, {\rm dex}$ and the number of galaxies in our low- and
high-$v_{\rm circ}$ samples.  In the case discussed by
\citet{dalcanton04}, we expect the low-$v_{\rm circ}$ galaxies to be
offset to lower $\Sigma_{\rm SFR}$ by $0.15 \, {\rm dex}$.  Therefore,
we cannot exclude offsets at the level expected by
\citet{dalcanton04}.

The star formation law assumed above is not likely correct given that
recent studies have found a strong molecular star formation law and no
atomic gas star formation law.  However, we can also determine the
expected $\Sigma_{\rm SFR}$ offset if the molecular fraction is set by
the mid-plane pressure and the molecular SFE is constant (see also
Section~\ref{sec:intro}).  We assume the same $\Sigma_{\rm HI+H_{2}}$
for a low- and high-$v_{\rm circ}$ galaxy, $R_{\rm mol} \propto P_{\rm
h}$, and $P_{\rm h} \propto \rho_{\rm HI+H_{2}} \sigma_{\rm gas}^{2}$,
where $\rho_{\rm HI+H_{2}}$ is the total hydrogen volume density and
the velocity dispersion of the gas is constant.  In this scenario,
$\rho_{\rm HI+H_{2}}$, $P_{\rm h}$, and $R_{\rm mol}$ are lower by a
factor of $\beta$ in the low-$v_{\rm circ}$ galaxy compared to the
high-$v_{\rm circ}$ galaxy.  The expected offset in $\Sigma_{\rm SFR}$
depends on $R_{\rm mol}$, which varies from 0.1 to 2.7 in our sample.
If $\beta = 2$ and $R_{\rm mol}$ of the high-$v_{\rm circ}$ galaxy is
0.1 (2.7), we expect $\Sigma_{\rm SFR}$ to be lower by $0.3 \, {\rm
dex}$ ($0.1 \, {\rm dex}$) in the low-$v_{\rm circ}$ galaxy compared
to the high-$v_{\rm circ}$ galaxy.  Given our uncertainties, this
level of offset would also be difficult to detect.  However, we can
reject our assumption that $P_{\rm h}$ is lower by a factor of two in
low-$v_{\rm circ}$ galaxies.  This is evident from
Section~\ref{sec:stability}, where we found no significant difference
in the mid-plane pressure distributions of the low- and high-$v_{\rm
circ}$ galaxies, even when considering only objects with the same
$\Sigma_{\rm HI+H_{2}}$.  Furthermore, we found no difference in the
$R_{\rm mol}$ distributions of the low- versus high-$v_{\rm circ}$
galaxies (Section~\ref{sec:no_trans}).

The two scenarios explored above predict offsets in intercept between
the low- and high-$v_{\rm circ}$ samples that are less than the
uncertainty.  Therefore, we cannot clearly exclude these options.
Nonetheless, our data show no evidence for a strong transition in SFE
at any circular velocity.  A simple interpretation that is consistent
with our data is that low-$v_{\rm circ}$ galaxies have a lower number
of molecular clouds per unit volume compared to high-$v_{\rm circ}$
galaxies at the same $\Sigma_{\rm HI+H_{2}}$, but lower only by the
ratio of the cold ISM scale heights in low- versus high-$v_{\rm circ}$
galaxies.  This results in the same total number of molecular clouds
within the beam for a low- and high-$v_{\rm circ}$ galaxy at the same
$\Sigma_{\rm HI+H_{2}}$ and thus gives the same $\Sigma_{\rm H_{2}}$
and $\Sigma_{\rm SFR}$ (assuming the molecular clouds have the same
density and that we average over evolutionary effects).  Note that the
above applies in the molecular-dominated regime.  We have few data
points below $\Sigma_{\rm HI+H_{2}} \sim 9 \, M_{\odot} \, {\rm
pc}^{-2}$, but expect that $\Sigma_{\rm SFR}$ can vary substantially
for galaxies with the same $\Sigma_{\rm HI+H_{2}}$, depending on the
physical processes that affect the molecular fraction (e.g., those
processes discussed in Krumholz et al. 2009 and Ostriker et al. 2010).

In conclusion, we interpret our result that there is no transition in
SFE at any circular velocity as evidence that scale height differences
at the level of about a factor of two do not significantly affect the
molecular fraction or SFE in bulgeless disk galaxies.  However,
offsets in SFE below our uncertainty level are still possible.

\subsection{Comparison to Star Formation Models}
\label{sec:models}
\citet{dalcanton04} very approximately estimated that the dust scale
heights of low-$v_{\rm circ}$ galaxies are about a factor of two
larger than high-$v_{\rm circ}$ galaxies.  Assuming our sample has a
similar range in scale height, our results indicate that these scale
height differences, which lead to gas volume density differences also
at the level of a factor of about two, do not lead to obvious
differences in the SFE.  Our results favor star formation models where
small-scale physical processes are more important than processes that
act on larger scales, of order the dust and cold gas scale height (10s
to $100 \, {\rm pc}$).  Based on their comparison to many star
formation models without an obvious favorite, \citet{leroy08}
discussed that physics below their resolution of $750 \, {\rm pc}$ is
likely most important for determining the SFE.  We contribute with a
further constraint that the SFE is likely affected primarily by
processes that act on scales smaller than the cold gas and dust scale
height.

We cannot exclude all star formation models that include large-scale
physics because there are processes that may affect star formation but
are neither affected by nor affect the scale height.  For example, our
sample has no power to constrain the effects of large-scale radial
processes, like shear \citep[see, e.g.,][]{hunter98}, on star
formation.  In addition, star formation may be affected by
environmental properties that depend on the gas volume density but
also depend on other variables that counteract a variable volume
density.  For example, the pressure in the ISM is related to the gas
volume density and velocity dispersion: $P \propto \rho_{\rm gas} \,
\sigma_{\rm gas}^2$.  While we expect the volume density to be lower
in low-$v_{\rm circ}$ galaxies, the velocity dispersion may be larger.
With the right combination of $\rho_{\rm gas}$ and $\sigma_{\rm gas}$,
there could be no difference in pressure between low- and high-$v_{\rm
circ}$ galaxies.  While the latter argument illustrates our
limitations, we note that there is currently no strong observational
reason to assume different gas velocity dispersions between the low-
and high-$v_{\rm circ}$ galaxies.  \citet{tamburro09} found some
variation in the central \ion{H}{1} velocity dispersion in eleven
galaxies ranging from early-type spirals to irregulars, but more
observations are needed.

Would leading star formation models have predicted a difference in SFE
in galaxies with scale heights that differ by about a factor of two?
In the \citet{krumholz09} model, $\Sigma_{\rm SFR}$ is a function of
metallicity, $\Sigma_{\rm gas}$, and the beam filling factor of $\sim
100 \, {\rm pc}$-sized atomic-molecular complexes.  There is no direct
dependence on the scale height, so we would not expect a transition in
SFE unless there is a transition in the metallicity or filling factor
with scale height.  We find no transition in oxygen abundance at any
circular velocity in our data, but there should be a correlation
between these two properties given the mass-metallicity relation
\citep[e.g.,][]{tremonti04}, which we also do not see.  The filling
factor of star-forming complexes is not well constrained, although
there is no {\it a priori} reason to suppose that it would be
different in low- versus high-$v_{\rm circ}$ objects.
\citet{krumholz09} do not predict a difference in SFE in galaxies with
different scale heights and the fact that we did not find a transition
in SFE at $v_{\rm circ} = 120 \, {\rm km \, s^{-1}}$ is not a strong
constraint on this model.

In the model where the mid-plane pressure sets the molecular fraction
and the molecular SFE is constant, the pressure is proportional to the
gas volume density, which we expect to vary between the low- and
high-$v_{\rm circ}$ galaxies.  If the gas velocity dispersion is
fixed, we would expect the low-$v_{\rm circ}$ galaxies to have lower
molecular to atomic surface density ratios and lower SFEs relative to
the high-$v_{\rm circ}$ galaxies .  However, we found no difference
between the mid-plane pressure, molecular to atomic surface density
ratio, or SFE distributions of the low- and high-$v_{\rm circ}$
galaxies. If our assumptions are correct, our result is inconsistent
with this model.  However, the offset in SFE expected for galaxies
with cold gas scale heights that differ by about a factor of two may
be less than our SFE uncertainties.  Furthermore, the
\citet{ostriker10} model relates the molecular fraction (or in their
terms, the fraction of gas in gravitationally bound complexes) to the
pressure of the diffuse component of the ISM.  We have a constraint
only on the scale height and volume density of the cold component of
the ISM; therefore the \citet{ostriker10} model may not predict
molecular fraction and SFE differences in our sample.  In general, our
results are somewhat more consistent with local models of star
formation, like the \citet{krumholz09} model, but we do not find
conclusive evidence for or against either the \citet{krumholz09} or
\citet{ostriker10} model.

One final matter to address is whether central measurements are
sufficient to determine if there is a transition in molecular
fraction, SFE, and/or stability at the dust structure transition of
$v_{\rm circ} = 120 \, {\rm km \, s^{-1}}$.  One might question the
use of central measurements because dust structure, SFE, and stability
may be affected by the higher gas and stellar densities and shorter
dynamical times characteristic of these regions.  The only approach
that will fully address this concern is to obtain off-center
measurements of the above properties.  Our single-beam CO data
currently limit us from carrying out this analysis.  Meanwhile, there
is some evidence that our central pointings are sufficient to address
these questions.  First, our measurements trace a significant fraction
of the disk: the $21\arcsec$ aperture probes physical scales of $0.7 -
3.2 \, {\rm kpc}$, which is similar to the disk scale lengths of our
sample ($0.7 - 3.4 \, {\rm kpc}$).  Second, the central regions of
bulgeless galaxies are more morphologically and kinematically similar
to the outskirts than in a galaxy with a bulge.  Finally, we expect a
galaxy to be less stable against gravitational collapse in the
center compared to the outer disk because the gas and stellar surface
densities are larger.  However, we find that both the low- and
high-$v_{\rm circ}$ galaxies are stable.  This suggests that there
would also not be a stability transition at $v_{\rm circ} = 120 \,
{\rm km \, s^{-1}}$ in off-center measurements because both the low-
and high-$v_{\rm circ}$ galaxies would be more stable.

\section{Summary}
We have presented a study of star formation in twenty
moderately-inclined, bulgeless disk galaxies. We found no transition
in star formation efficiency ($\Sigma_{\rm SFR}/\Sigma_{\rm
HI+H_{2}}$) or disk stability at $v_{\rm circ} = 120 \, {\rm km \,
s^{-1}}$.  This circular velocity was previously found to be
associated with a transition in the vertical dust structure of
edge-on, bulgeless disk galaxies that is most likely due to a
transition in the scale height of the cold ISM.  We also found no
transition in star formation efficiency or disk stability at any
circular velocity probed by our sample.  Our results demonstrate that
the scale height of the cold ISM does not play a major role in setting
the molecular fraction or the star formation efficiency.  We also
found decreasing star formation efficiency with lower oxygen
abundance, which we estimated from the stellar mass. This result is
consistent with the recent \citet{krumholz09} model, but a sample with
a large range of metallicities within a small range of gas surface
density would provide a more constraining test of the model.  In
general, our results are most consistent with local models of star
formation that include physical processes that act on smaller scales
than the dust and cold gas scale height (10s to $100 \, {\rm pc}$).

\acknowledgements We thank Frank Bigiel for providing us with his
data.  We are also grateful to Todd A. Thompson for helpful comments
and discussion, Richard W. Pogge for supplying the narrowband filters
used in the H$\alpha$ observations, Roberto J. Assef, David W. Atlee,
and Katharine J. Schlesinger for obtaining some of the H$\alpha$
observations, and the referee for comments that improved this work.
L.C.W. gratefully acknowledges support from an NSF Graduate Research
Fellowship and an Ohio State University Distinguished University
Fellowship.  PM is grateful for support from the NSF via award
AST-0705170.  U.L. acknowledges financial support from the research
projects AYA2007-67625-C02-02 and AYA2011-24728 from the Spanish
Ministerio de Ciencia y Educaci\'on and from the Junta de Andaluc\'\i
a.  This work is based in part on observations made with the Spitzer
Space Telescope, which is operated by the Jet Propulsion Laboratory,
California Institute of Technology under a contract with NASA. Support
for this work was provided by NASA through an award issued by
JPL/Caltech.  This publication makes use of data products from the Two
Micron All Sky Survey, which is a joint project of the University of
Massachusetts and the Infrared Processing and Analysis
Center/California Institute of Technology, funded by the National
Aeronautics and Space Administration and the National Science
Foundation.

\clearpage
\begin{landscape}
\begin{deluxetable}{lcccccccccccc}
\setlength{\tabcolsep}{1pt}
\tablewidth{0pt}
\tabletypesize{\scriptsize}
\tablecaption{General Galaxy Properties}
\tablehead{
\colhead{Source} &
\colhead{RA} &
\colhead{DEC} &
\colhead{D} &
\colhead{$D_{25}$} &
\colhead{$B$} &
\colhead{$M_{B}$} &
\colhead{Type} &
\colhead{$V_{\rm sys}$} &
\colhead{$v_{\rm circ}$} &
\colhead{PA} &
\colhead{{\it i}} &
\colhead{$W_{20}$} \\
\colhead{} &
\colhead{(hh:mm:ss.s)} &
\colhead{(dd:mm:ss)} &
\colhead{(Mpc)} &
\colhead{(arcsec)} &
\colhead{(mag)} &
\colhead{(mag)} &
\colhead{} &
\colhead{(${\rm km\, s^{-1}}$)} &
\colhead{(${\rm km\, s^{-1}}$)} &
\colhead{($\degr$)} &
\colhead{($\degr$)} &
\colhead{(${\rm km \, s^{-1}}$)} \\
\colhead{(1)} &
\colhead{(2)} &
\colhead{(3)} &
\colhead{(4)} &
\colhead{(5)} &
\colhead{(6)} &
\colhead{(7)} &
\colhead{(8)} &
\colhead{(9)} &
\colhead{(10)} &
\colhead{(11)} &
\colhead{(12)} &
\colhead{(13)}
}
\startdata

NGC~0337      &  00:59:50.0   &   -07:34:41    &  20.7 [T88]   &  173  &  11.44  & -20.14 & 7.0  &  $1646   \pm 2   $  &  $145    ^{+5    }_{-4    } $ &   $118    \pm 5    $ &  $44    \pm  2    $ &  261   \\
PGC~3853      &  01:05:04.8   &   -06:12:46    &  11.4 [T08]   &  250  &  11.98  & -18.30 & 7.0  &  $1094.7 \pm 0.4 $  &  $128.1  ^{+1.6  }_{-2    } $ &   $105.3  \pm 0.2  $ &  $41.4  \pm  1.1  $ &  192   \\
PGC~6667      &  01:49:10.3   &   -10:03:45    &  24.6 [T88]   &  173  &  12.92  & -19.03 & 6.7  &  $1989.2 \pm 0.6 $  &  $155    ^{+4    }_{-4    } $ &   $122.9  \pm 1.8  $ &  $34.0  \pm  1.1  $ &  198   \\
ESO~544-G030  &  02:14:57.2   &   -20:12:40    &  13.9 [T08]   &  123  &  13.25  & -17.47 & 7.7  &  $1608.4 \pm 1.0$   &  $100.9  ^{+1.4  }_{-2    } $ &   $107.6  \pm 1.1  $ &  $48.5  \pm  1.2  $ &  146   \\
UGC~1862      &  02:24:24.8   &   -02:09:41    &  22.3 [T08]   &  99.6 &  13.47  & -18.27 & 7.0  &  $1382.9 \pm 0.4 $  &  $55     ^{+6    }_{-5    } $ &   $21.7   \pm 1.7  $ &  $43    \pm  4    $ &  125   \\
ESO~418-G008  &  03:31:30.8   &   -30:12:46    &  23.6 [T08]   &  70.5 &  13.65  & -18.21 & 8.0  &  $1195.4 \pm 0.3 $  &  $74.1   ^{+0.6  }_{-0.6  } $ &   $317.9  \pm 1.1  $ &  $55.6  \pm  1.4  $ &  140   \\
ESO~555-G027  &  06:03:36.6   &   -20:39:17    &  24.3 [T88]   &  138  &  13.18  & -18.75 & 7.0  &  $1978.7 \pm 0.4 $  &  $190    ^{+40   }_{-30   } $ &   $221.5  \pm 0.3  $ &  $21    \pm  4    $ &  162   \\
NGC~2805      &  09:20:20.4   &   +64:06:12    &  28.0 [T88]   &  379  &  11.17  & -21.07 & 7.0  &  $1732.6 \pm 0.6 $  &  $81     ^{+7    }_{-8    } $ &   $300    \pm 3    $ &  $38    \pm  4    $ &  120   \\
ESO~501-G023  &  10:35:23.6   &   -24:45:21    &  7.01 [T08]   &  208  &  12.86  & -16.37 & 8.0  &  $1046.8 \pm 0.7 $  &  $46     ^{+18   }_{-8    } $ &   $224    \pm 2    $ &  $37    \pm  12   $ &  83    \\
UGC~6446      &  11:26:40.6   &   +53:44:58    &  18.0 [T08]   &  213  &  13.30  & -17.98 & 7.0  &  $645.5  \pm 0.6 $  &  $79.7   ^{+1.3  }_{-0.8  } $ &   $189.4  \pm 0.5  $ &  $52.5  \pm  1.9  $ &  150   \\
NGC~3794      &  11:40:54.8   &   +56:12:10    &  19.2 [T08]   &  134  &  13.23  & -18.19 & 6.5  &  $1384.9 \pm 0.7 $  &  $103.3  ^{+1.1  }_{-1.0  } $ &   $123.1  \pm 1.1  $ &  $54.8  \pm  1.3  $ &  182   \\
NGC~3906      &  11:49:40.2   &   +48:25:30    &  18.3 [...]   &  112  &  13.50  & -17.81 & 7.0  &  $959.44 \pm 0.7 $  &  $65     ^{+30   }_{-16   } $ &   $180    \pm 20   $ &  $16    \pm  5    $ &  49    \\
UGC~6930      &  11:57:17.2   &   +49:17:08    &  17.0 [T88]   &  262  &  12.38  & -18.77 & 7.0  &  $776.7  \pm 0.7 $  &  $121    ^{+20   }_{-15   } $ &   $39.5   \pm 0.5  $ &  $25    \pm  4    $ &  140   \\
NGC~4519      &  12:33:30.5   &   +08:39:16    &  19.6 [T08]   &  190  &  12.15  & -19.31 & 7.0  &  $1218.1 \pm 1.0 $  &  $112    ^{+8    }_{-7    } $ &   $355    \pm 2    $ &  $42    \pm  3    $ &  218   \\
NGC~4561      &  12:36:08.6   &   +19:19:26    &  12.3 [T88]   &  90.8 &  12.82  & -17.63 & 8.0  &  $1402.2 \pm 0.9 $  &  $57     ^{+6    }_{-5    } $ &   $227    \pm 8    $ &  $34    \pm  4    $ &  171   \\
NGC~4713      &  12:49:58.1   &   +05:18:39    &  14.9 [T08]   &  162  &  11.85  & -19.02 & 7.0  &  $654.5  \pm 0.5 $  &  $110.9  ^{+2    }_{-1.8  } $ &   $274.0  \pm 0.6  $ &  $45.2  \pm  1.2  $ &  176   \\
NGC~4942      &  13:04:19.2   &   -07:39:00    &  28.5 [T88]   &  112  &  13.27  & -19.00 & 7.0  &  $1741   \pm 2   $  &  $124    ^{+6    }_{-5    } $ &   $137.3  \pm 0.8  $ &  $37    \pm  2    $ &  177   \\
NGC~5964      &  15:37:36.3   &   +05:58:28    &  24.7 [T88]   &  250  &  12.28  & -19.68 & 7.0  &  $1447.1 \pm 1.2 $  &  $168    ^{+18   }_{-14   } $ &   $136.7  \pm 1.2  $ &  $32    \pm  3    $ &  208   \\
NGC~6509      &  17:59:24.9   &   +06:17:12    &  28.2 [T88]   &  95.1 &  12.12  & -20.13 & 7.0  &  $1811.0 \pm 0.4 $  &  $153    ^{+12   }_{-9    } $ &   $280.8  \pm 1.1  $ &  $41    \pm  4    $ &  266   \\
IC~1291	      &  18:33:51.5   &   +49:16:45    &  31.5 [T88]   &  109  &  13.28  & -19.21 & 8.0  &  $1951.0 \pm 1.1 $  &  $189    ^{+17   }_{-14   } $ &   $131    \pm 2    $ &  $28    \pm  3    $ &  209   \\

\enddata

\tablecomments{Column 1: Object name; Column 2 and 3: Right ascension
  and declination (J2000.0) from \citet{RC3}; Column 4: Distance and
  distance reference.  Distances are derived using the Tully-Fisher
  relation, except for NGC~3906.  T08: \citet{tully08}, T88:
  \citet{tully88}, and the NGC~3906 distance is from the \citet{RC3}
  heliocentric velocity, corrected for Virgo infall using
  \citet{mould00} and using $H_{0} = 71 \, {\rm km \, s^{-1} \,
  Mpc^{-1}}$.  Column 5: Major isophotal diameter at $25 \, {\rm mag
  \, arcsec^{-2}}$ in the B band, from \citet{RC3}.  Column 6:
  Apparent blue magnitude, corrected for Galactic and internal
  extinction and redshift.  Values are from \citet{RC3}, except for
  NGC~4942 and PGC~6667, which are from \citet{doyle05} and are only
  corrected for Galactic extinction.  Column 7: Absolute blue
  magnitude, calculated from the apparent magnitude in Column 6 and
  the distance in Column 4.  Column 8: Morphological type from
  \citet{RC3}.  Column 9: Systemic velocity, corrected to the
  heliocentric reference frame. Column 10: Circular velocity.  Columns
  9 and 10 were derived from the VLA \ion{H}{1} rotation curve
  analysis in Paper~I.  Column 11: Position angle of the major axis
  (degrees N to E to receding side).  Column 12: Inclination.  Columns
  11 and 12 were derived from a combination of the rotation curve
  analyses on the \ion{H}{1} data and ellipse fits of the IRAC $3.6 \,
  \mu {\rm m}$ data.  Columns 9-12 were originally presented in
  Table~6 of Paper~I.  Column 13: Width of the HI line at 20\% of the
  peak flux density, corrected for the spectral resolution but not
  turbulent broadening.  The uncertainty in $W_{20}$ is $5\, {\rm km
  \, s^{-1}}$ for all objects except UGC~6446 and NGC~3906, where the
  uncertainty is $3 \, {\rm km \, s^{-1}}$ and $10\, {\rm km \,
  s^{-1}}$, respectively.  Column~13 was originally presented in
  Table~4 of Paper~I.}

\label{tab:sample}
\end{deluxetable}
\clearpage
\end{landscape}

\clearpage

\clearpage
\begin{landscape}
\begin{deluxetable}{lcccccccccccc}
\setlength{\tabcolsep}{1pt}
\tablewidth{0pt}
\tabletypesize{\scriptsize}
\tablecaption{Measured and Literature Properties}
\tablehead{
\colhead{Source} &
\colhead{RA} &
\colhead{DEC} &
\colhead{$F_{\rm H\alpha}$} &
\colhead{$F_{\rm PAH}$} &
\colhead{$\langle I_{\rm HI} \rangle$} &
\colhead{$B_{\rm maj}$} &
\colhead{$B_{\rm min}$} &
\colhead{$I_{\rm CO}$} &
\colhead{$F_{4.5}^{21\arcsec}$} &
\colhead{$F_{4.5}^{D25}$} &
\colhead{$J$} &
\colhead{$K_{\rm s}$} \\
\colhead{} &
\colhead{(hh:mm:ss.s)} &
\colhead{(dd:mm:ss)} &
\colhead{($10^{-14} \, {\rm erg \, s^{-1} \, cm^{-2}}$)} &
\colhead{(mJy)} &
\colhead{(${\rm Jy \, beam^{-1} \, km \, s^{-1}}$)} &
\colhead{(arcsec)} &
\colhead{(arcsec)} &
\colhead{(${\rm K \, km\, s^{-1}}$)} &
\colhead{(mJy)} &
\colhead{(mJy)} &
\colhead{(mag)} &
\colhead{(mag)} \\
\colhead{(1)} &
\colhead{(2)} &
\colhead{(3)} &
\colhead{(4)} &
\colhead{(5)} &
\colhead{(6)} &
\colhead{(7)} &
\colhead{(8)} &
\colhead{(9)} &
\colhead{(10)} &
\colhead{(11)} &
\colhead{(12)} &
\colhead{(13)}
}
\startdata
NGC~0337      &  00:59:49.9	& -07:34:44	& $ 10.2	 \pm   0.05	$    &  $ 73	\pm  8    $ &  $1.28   \pm      0.17   $  &     25.77 & 15.06  & $ 4.7    \pm 0.3  $ & $ 11.7  \pm  1.2	$& $64     \pm 6    $ & $ 9.876  \pm 0.025 $& $ 9.059  \pm 0.045$	  \\
PGC~3853      &  01:05:04.9	& -06:12:45	& $ 0.98	 \pm   0.04	$    &  $ 6.9	\pm  0.8  $ &  $0.36   \pm      0.05   $  &     25.33 & 15.47  & $ 1.39   \pm 0.18 $ & $ 3.4   \pm  0.3	$& $44     \pm 4    $ & $ 10.031 \pm 0.037 $& $ 9.280  \pm 0.080$	  \\
PGC~6667      &  01:49:10.3	& -10:03:40	& $ 2.37	 \pm   0.02	$    &  $ 8.5	\pm  0.9  $ &  $0.90   \pm      0.12   $  &     26.00 & 21.00  & $ 0.71   \pm 0.13 $ & $ 2.4   \pm  0.2	$& $15.2   \pm 1.5  $ & $ 11.678 \pm 0.037 $& $ 10.951 \pm 0.080$	  \\
ESO~544-G030  &  02:14:56.8	& -20:12:44	& $ 0.85	 \pm   0.02	$    &  $ 5.3	\pm  0.6  $ &  $0.59   \pm      0.08   $  &     27.78 & 12.01  & $ <0.4		$ & $ 2.5   \pm  0.3	$& $12.8   \pm 1.3  $ & $ ...		   $& $ ...      $	  \\
UGC~1862      &  02:24:24.8	& -02:09:44	& $ 1.97	 \pm   0.02	$    &  $ 4.2	\pm  0.5  $ &  $0.24   \pm      0.03   $  &     23.97 & 15.32  & $ 0.60   \pm 0.10 $ & $ 2.2   \pm  0.2	$& $12.4   \pm 1.2  $ & $ 11.927 \pm 0.023 $& $ 11.177 \pm 0.054$	  \\
ESO~418-G008  &  03:31:30.7	& -30:12:48	& $ 1.84	 \pm   0.03	$    &  $ 6.7	\pm  0.8  $ &  $0.92   \pm      0.12   $  &     32.00 & 21.00  & $ <0.9            $ & $ 2.7   \pm  0.3	$& $6.7    \pm 0.7  $ & $ 12.752 \pm 0.049 $& $ 12.169 \pm 0.126$	  \\
ESO~555-G027  &  06:03:36.8	& -20:39:15	& $ 1.42	 \pm   0.05	$    &  $ 9.3	\pm  1.0  $ &  $0.44   \pm      0.06   $  &     29.00 & 21.00  & $        ...	$ & $ 2.4   \pm  0.2	$& $16.1   \pm 1.6  $ & $ 11.984 \pm 0.045 $& $ 11.271 \pm 0.103$	  \\
NGC~2805      &  09:20:20.3	& +64:06:11	& $ 2.53	 \pm   0.017    $    &  $ 15.5  \pm  1.7  $ &  $0.34   \pm      0.04   $  &     21.00 & 21.00  & $ 2.34   \pm 0.10 $ & $ 4.2   \pm  0.4	$& $41     \pm 4    $ & $ 10.827 \pm 0.026 $& $ 10.117 \pm 0.046$	  \\
ESO~501-G023  &  10:35:23.3	& -24:45:15	& $ 0.76	 \pm   0.03	$    &  $ 1.04  \pm  0.14 $ &  $0.134   \pm     0.017  $  &     30.30 & 12.51  & $ <0.6		$ & $ 0.90  \pm  0.09	$& $9.9    \pm 1.0  $ & $ ...		   $& $ ...      $	 \\
UGC~6446      &  11:26:40.4	& +53:44:48	& $ 1.19	 \pm   0.02	$    &  $ 0.44  \pm  0.09 $ &  $0.43   \pm      0.06   $  &     21.00 & 21.00  & $ <0.4		$ & $ 0.97  \pm  0.10	$& $9.5    \pm 1.0  $ & $ ...		   $& $ ...      $	 \\
NGC~3794      &  11:40:54.3	& +56:12:07	& $ 2.37	 \pm   0.016    $    &  $ 7.3	\pm  0.9  $ &  $0.49   \pm      0.06   $  &     21.00 & 21.00  & $ 0.72   \pm 0.15 $ & $ 3.5   \pm  0.4	$& $13.0   \pm 1.3  $ & $ 11.871 \pm 0.039 $& $ 11.005 \pm 0.072$	  \\
NGC~3906      &  11:49:39.9	& +48:25:32	& $ 2.25	 \pm   0.02	$    &  $ 7.7	\pm  0.9  $ &  $0.25   \pm      0.03   $  &     21.00 & 21.00  & $ 0.37   \pm 0.12 $ & $ 2.8   \pm  0.3	$& $14.0   \pm 1.4  $ & $ 11.788 \pm 0.045 $& $ 11.007 \pm 0.067$	  \\
UGC~6930      &  11:57:17.4	& +49:16:58	& $ 0.76	 \pm   0.03	$    &  $ 6.8	\pm  0.8  $ &  $0.25   \pm      0.03   $  &     21.00 & 21.00  & $ 1.31   \pm 0.19 $ & $ 2.6   \pm  0.3	$& $29     \pm 3    $ & $ 11.752 \pm 0.037 $& $ 11.153 \pm 0.068$	  \\
NGC~4519      &  12:33:30.3	& +08:39:18	& $ 3.0	         \pm   0.03	$    &  $ 44	\pm  5    $ &  $0.99   \pm      0.13   $  &     51.91 & 18.77  & $ 2.9    \pm 0.3  $ & $ 6.9   \pm  0.7	$& $36     \pm 4    $ & $ 10.499 \pm 0.037 $& $ 9.548  \pm 0.059$	  \\
NGC~4561      &  12:36:08.2	& +19:19:22	& $ 4.55	 \pm   0.03	$    &  $ 7.9	\pm  1.0  $ &  $1.02   \pm      0.13   $  &     21.00 & 21.00  & $ 0.68   \pm 0.12 $ & $ 4.1   \pm  0.4	$& $14.8   \pm 1.5  $ & $ 11.480 \pm 0.046 $& $ 10.617 \pm 0.074$	  \\
NGC~4713      &  12:49:58.0	& +05:18:41	& $ 5.35	 \pm   0.03	$    &  $ 35	\pm  4    $ &  $0.69   \pm      0.09   $  &     21.00 & 21.00  & $ 4.5    \pm 0.2  $ & $ 7.1   \pm  0.7	$& $44     \pm 4    $ & $ 10.367 \pm 0.026 $& $ 9.737  \pm 0.053$	  \\
NGC~4942      &  13:04:19.1	& -07:38:58	& $ 3.55	 \pm   0.04	$    &  $ 10.1  \pm  1.1  $ &  $0.49   \pm      0.06   $  &     22.00 & 21.00  & $ 1.4    \pm 0.2  $ & $ 3.4   \pm  0.3	$& $12.9   \pm 1.3  $ & $ 11.617 \pm 0.058 $& $ 10.428 \pm 0.092$	  \\
NGC~5964      &  15:37:36.2	& +05:58:27	& $ 2.11	 \pm   0.05	$    &  $ 9.0	\pm  1.0  $ &  $0.30   \pm      0.04   $  &     21.00 & 21.00  & $ 0.89   \pm 0.17 $ & $ 3.4   \pm  0.3	$& $40     \pm 4    $ & $ 12.384 \pm 0.058 $& $ 11.794 \pm 0.113$	  \\
NGC~6509      &  17:59:25.2	& +06:17:11	& $ 4.62	 \pm   0.03	$    &  $ 41	\pm  5    $ &  $0.50   \pm      0.07   $  &     21.00 & 21.00  & $ 6.0    \pm 0.2  $ & $ 10.6  \pm  1.1	$& $42     \pm 4    $ & $ 10.284 \pm 0.024 $& $ 9.663  \pm 0.043$	  \\
IC~1291	      &  18:33:52.5	& +49:16:42	& $ 6.02	 \pm   0.02	$    &  $ 9.3	\pm  1.0  $ &  $0.70   \pm      0.09   $  &     21.00 & 21.00  & $ 0.37   \pm 0.08 $ & $ 2.4   \pm  0.2	$& $10.2   \pm 1.0  $ & $ 13.125 \pm 0.052 $& $ 12.689 \pm 0.152$	  \\
\enddata

\tablecomments{Column 1: Object name.  Column 2 and 3: RA and DEC
  (J2000.0) of the pointing center of the IRAM 30m CO(1-0)
  observations.  All measurements within a 21$\arcsec$-diameter
  circular aperture are centered on these coordinates.  Column 4:
  H$\alpha$ flux within a 21$\arcsec$-diameter circular aperture,
  corrected for Galactic extinction and \ion{N}{2} emission. We quote
  only measurement uncertainties here.  Column 5: PAH flux density
  within a 21$\arcsec$-diameter circular aperture.  Column 6: Average
  integrated \ion{H}{1} line intensity within a 21$\arcsec$-diameter
  circular aperture, measured from image convolved to have beam major
  and minor axes given in Column 7 and 8. Column 7 and 8: Beam major
  and minor axes of image used for HI line intensity measurement.
  Column 9: Integrated CO(1-0) line intensity.  Non-detections are
  quoted as $3 \sigma$ upper limits. Column 10: 4.5$\mu {\rm m}$ flux
  density within a 21$\arcsec$-diameter circular aperture.  Column 11:
  Total 4.5$\mu {\rm m}$ flux density. Column 12: Total $J$-band
  magnitude from 2MASS, corrected for Galactic extinction.  Column 13:
  Total $K_{\rm s}$-band magnitude from 2MASS, corrected for Galactic
  extinction.  }
\label{tab:obs_prop}
\end{deluxetable}
\clearpage
\end{landscape}
\clearpage

\begin{deluxetable}{lcccc}
\tablecolumns{5}
\tablewidth{0pt}
\tabletypesize{\scriptsize}
\tablecaption{H$\alpha$ Observations} 
\tablehead{
\colhead{Galaxy} &
\colhead{$t_{663nb15}$ (min)} &
\colhead{Run(s)} &
\colhead{$t_{693nb15}$ (min)} &
\colhead{Run(s)} \\
\colhead{(1)} &
\colhead{(2)} &
\colhead{(3)} &
\colhead{(4)} &
\colhead{(5)}
}
\startdata
NGC0337   	& 2400  & 3    & 2400  & 3     \\
PGC3853   	& 1800  & 4    & 1800  & 1     \\
PGC6667   	& 4500  & 1,3  & 4500  & 1,3   \\
ESO544-G030   	& 5400  & 1,3  & 5400  & 1,3   \\
UGC1862   	& 4500  & 1,3  & 5400  & 1,3   \\
ESO418-G008   	& 2700  & 1    & 2700  & 1     \\
ESO555-G027   	& 2700  & 1    & 2700  & 1     \\
NGC2805   	& 2400  & 3    & 2400  & 3     \\
ESO501-G023   	& 3600  & 1    & 3600  & 1     \\
UGC6446   	& 3900  & 2    & 4200  & 2     \\
NGC3794   	& 6000  & 1,3  & 6000  & 1,3   \\
NGC3906   	& 3200  & 1,2,3& 3000  & 1,2   \\
UGC6930   	& 3600  & 2    & 3600  & 2     \\
NGC4519   	& 3600  & 2    & 3600  & 2     \\
NGC4561   	& 3600  & 1    & 3600  & 1     \\
NGC4713   	& 2400  & 1,2  & 2400  & 1,2   \\
NGC4942   	& 4500  & 1,2  & 3600  & 2     \\
NGC5964   	& 3600  & 2    & 3600  & 2     \\
NGC6509   	& 4800  & 2    & 3600  & 2     \\
IC1291   	& 3600  & 2    & 3600  & 2     \\
\enddata

\tablecomments{Column 1: Object name. Column 2: Total exposure time in
the 663nb15 filter. Column 3: Observing run code.  Column 4: Total
exposure time in the 693nb15 filter. Column 5: Observing run code.
Observing run codes are: (1) Jan 2007; (2) May/June 2007; (3) Nov
2007; (4) Jan 2008.  }
\label{tbl:halphalog}
\end{deluxetable}

\begin{deluxetable}{lccccccccccccc}
\setlength{\tabcolsep}{1pt}
\tablewidth{0pt}
\tabletypesize{\scriptsize}
\tablecaption{Derived Properties}
\tablehead{
\colhead{Source} &
\colhead{$R_{21\arcsec}$} &
\colhead{${\rm log} \, \Sigma_{\rm HI}$} &
\colhead{${\rm log} \, \Sigma_{\rm H2}$} &
\colhead{${\rm log} \, \Sigma_{\rm SFR}$} &
\colhead{${\rm log} \, \Sigma_{\ast}$} &
\colhead{${\rm log} \, M_{\ast}$} &
\colhead{$12 + {\rm log}(O/H)$} &
\colhead{$l_{\ast}$} &
\colhead{$\kappa$} &
\colhead{$Q_{\rm gas}$} &
\colhead{$Q_{\ast}$} &
\colhead{$Q_{\rm gas+stars}$} &
\colhead{${\rm log} \, P_{\rm h}$} \\
\colhead{(1)} &
\colhead{(2)} &
\colhead{(3)} &
\colhead{(4)} &
\colhead{(5)} &
\colhead{(6)} &
\colhead{(7)} &
\colhead{(8)} &
\colhead{(9)} &
\colhead{(10)} &
\colhead{(11)} &
\colhead{(12)} &
\colhead{(13)} &
\colhead{(14)} 
}
\startdata
NGC~0337     &   2.11  & 	1.33  & 	1.18	 &  -1.67  &   2.49  & 	 9.9   & 	 8.93  & 	 1.7   & 	0.17  & 	2.7  & 	2.9  & 2.0     & 5.32	 \\
PGC~3853     &   1.16  & 	0.78  & 	0.67	 &  -2.68  &   1.88  & 	 9.1   & 	 8.68  & 	 1.6   & 	0.22  & 	12   & 	8    & 5       & 4.42	 \\
PGC~6667     &   2.51  & 	1.09  & 	0.42	 &  -2.44  &   1.73  & 	 9.3   & 	 8.75  & 	 2.4   & 	0.12  & 	4.8  & 	6    & 3.2     & 4.50	 \\
ESO~544-G030 &   1.42  & 	1.02  & 	$<0.09$	 &  -2.83  &   1.51  & 	 8.6   & 	 8.46  & 	 1.1   & 	0.12  & 	6.0  & 	5    & 2.9     & 4.39	 \\
UGC~1862     &   2.27  & 	0.62  & 	0.29	 &  -2.70  &   1.67  & 	 9.2   & 	 8.70  & 	 1.5   & 	0.07  & 	7    & 	3.0  & 2.3     & 4.04	 \\
ESO~418-G008 &   2.40  & 	0.84  &         $<0.34$	 &  -2.72  &   1.40  & 	 8.7   & 	 8.50  & 	 1.1   & 	0.07  & 	4.5  & 	3    & 2.0     & 4.21	 \\
ESO~555-G027 &   2.47  & 	0.78  & 	...	 &  -2.44  &   1.77  & 	 9.3   & 	 8.74  & 	 1.9   & 	0.12  & 	...  &  5    & ...     & ... 	 \\
NGC~2805     &   2.85  & 	0.74  & 	0.92	 &  -2.28  &   1.93  & 	 9.8   & 	 8.91  & 	 2.3   & 	0.041 & 	1.8  & 	1.6  & 1.1     & 4.52	 \\
ESO~501-G023 &   0.71  & 	0.40  &       $<0.32$  	 &  -3.16  &   1.01  & 	 7.8   & 	 8.00  & 	 0.72  & 	0.12  & 	16   & 	8    & 5       & 3.77	 \\
UGC~6446     &   1.83  & 	0.73  &       $<-0.008$	 &  -3.21  &   1.01  & 	 8.6   & 	 8.47  & 	 1.1   & 	0.07  & 	6.5  & 	5    & 2.9     & 3.88	 \\
NGC~3794     &   1.96  & 	0.75  & 	0.27	 &  -2.64  &   1.93  & 	 9.2   & 	 8.72  & 	 1.5   & 	0.15  & 	12   & 	5    & 3.9     & 4.26	 \\
NGC~3906     &   1.86  & 	0.69  & 	0.20	 &  -2.42  &   1.94  & 	 9.1   & 	 8.67  & 	 0.93  & 	0.09  & 	8    & 	2.1  & 1.8     & 4.27	 \\
UGC~6930     &   1.73  & 	0.66  & 	0.72	 &  -2.63  &   1.62  & 	 9.1   & 	 8.66  & 	 2.1   & 	0.18  & 	11   & 	10   & 6       & 4.24	 \\
NGC~4519     &   2.00  & 	0.82  & 	0.99	 &  -1.93  &   2.46  & 	 9.8   & 	 8.90  & 	 2.0   & 	0.22  & 	8    & 	4    & 3.7     & 4.83	 \\
NGC~4561     &   1.25  & 	1.23  & 	0.40	 &  -2.32  &   2.15  & 	 8.9   & 	 8.58  & 	 0.69  & 	0.14  &  	4.2  & 	2.3  & 1.7     & 4.98	 \\
NGC~4713     &   1.52  & 	0.99  & 	1.15	 &  -1.98  &   1.99  & 	 9.2   & 	 8.70  & 	 1.4   & 	0.25  & 	6.3  & 	7    & 4.0     & 4.94	 \\
NGC~4942     &   2.90  & 	0.88  & 	0.69	 &  -2.35  &   2.52  & 	 10.0  & 	 8.96  & 	 1.8   & 	0.12  & 	5.8  & 	2.1  & 1.9     & 4.72	 \\
NGC~5964     &   2.52  & 	0.71  & 	0.53	 &  -2.44  &   1.70  & 	 9.5   & 	 8.82  & 	 3.4   & 	0.055 & 	3.9  & 	3    & 2.2     & 4.11	 \\
NGC~6509     &   2.87  & 	0.89  & 	1.31	 &  -1.92  &   2.18  & 	 9.7   & 	 8.88  & 	 1.7   & 	0.22  & 	4.7  & 	6    & 3.3     & 5.06	 \\
IC~1291	     &   3.21  & 	1.10  & 	0.17	 &  -2.19  &   1.35  & 	 8.9   & 	 8.60  & 	 2.1   & 	0.18  & 	8    & 	13   & 5       & 4.37	 \\
\enddata

\tablecomments{Column 1: Object name. Column 2: Physical size of
$21\arcsec$ (kpc). All surface density measurements are within a
$21\arcsec$-diameter aperture. Column 3: HI mass surface density
($M_{\odot} \, {\rm pc^{-2}}$).  The typical uncertainty is $0.06 \,
{\rm dex}$.  Column 4: ${\rm H_{2}}$ mass surface density ($M_{\odot}
\, {\rm pc^{-2}}$), computed with $X_{\rm CO} = 2.8 \times 10^{20} \,
{\rm cm^{-2} \, (K \, km \, s^{-1})^{-1}}$. The typical uncertainty is
$0.07 \, {\rm dex}$.  Column 5: SFR surface density, computed assuming
the Kroupa-type IMF from \citet{calzetti07} ($M_{\odot} \, {\rm
yr^{-1} \, kpc^{-2}}$).  We assume a typical uncertainty of $0.2 \,
{\rm dex}$.  Column 6: Stellar mass surface density ($M_{\odot} \,
{\rm pc^{-2}}$).  We assign a a typical uncertainty of $0.2 \, {\rm
dex}$.  Column 7: Total stellar mass ($M_{\odot}$).  We assign typical
uncertainty of $0.3 \, {\rm dex}$.  Column 8: Oxygen abundance derived
from the total stellar mass and the mass-metallicity relation of
\citet{tremonti04}.  We derive a typical statistical uncertainty of
$0.11 \, {\rm dex}$, but note that strong-line metallicity methods
return $12+{\rm log}(O/H)$ values as disparate as $0.5 \, {\rm dex}$.
Column 9: Stellar scale length (kpc).  We estimate the uncertainty to
be 20\%.  Column 10: Epicyclic frequency (km/s/pc), evaluated at
$5.25\arcsec$.  We estimate the uncertainty to be 30\%.  Column 11:
Gas stability parameter. The typical uncertainty is 30\%. Column 12:
Stellar stability parameter.  The typical uncertainty is 60\%.  Column
13: Combined gas and stellar stability parameter.  The typical
uncertainty is 40\%.  Column 14: Mid-plane pressure (${\rm K \,
cm^{-3}}$).  The typical uncertainty is $0.17 \, {\rm dex}$.}
\label{tab:der_prop}
\end{deluxetable}

\end{document}